%% file: main.tex
\documentclass[aps,prd,reprint,nofootinbib]{revtex4-2}
\input{config/packages}

\begin{document}

\input{config/macros}

\graphicspath{{figures/}}

\title{Geometric noise spectrum in interferometers}
\author{Laurent Freidel}
\affiliation{Perimeter Institute for Theoretical Physics, 31 Caroline St. North, Waterloo, ON, N2L 2Y5, Canada}
\author{Robin Oberfrank}
\affiliation{Perimeter Institute for Theoretical Physics, 31 Caroline St. North, Waterloo, ON, N2L 2Y5, Canada}
\affiliation{Department of Physics and Astronomy, University of Waterloo, Waterloo, ON, N2L 3G1, Canada}

\begin{abstract}
\setlength{\leftskip}{1em}
\setlength{\rightskip}{1em}
	We study the power spectral density of time delay fluctuations in an
	interferometer as a potential low-energy quantum gravitational observable. We
	derive a general expression for the spectrum in terms of the Wightman function
	of linear metric perturbations, which we then apply to a variety of cases. We
	analyze the intrinsic graviton fluctuations in the vacuum, thermal, and
	squeezed states, as well as the fluctuations induced by the vacuum
	stress-energy of a massless scalar field. We find that the resulting spectra
	are free of ultraviolet divergences and that, while thermal and squeezed
	states provide a natural amplification mechanism, the spectra remain
	suppressed by the Planck scale.
\end{abstract}

\maketitle

\newpage
\setcounter{page}{1}
\tableofcontents


\newcommand{\cancel}[1]{\begingroup\color{gray}#1\endgroup}
\newcommand{\new}[1]{\begingroup\color{red}#1\endgroup}
\newpage
\input{sections/introduction}
\input{sections/sec2_spectrum}

\input{sections/sec3_intrinsic}

\input{sections/sec4_induced}
\input{sections/conclusion}

\section*{Acknowledgements}

We would like to thank Simone Speziale for fruitful
conversations. Robin would like to thank Jacqueline Caminiti for a useful discussion and Antonia Seifert, Kiana Salehi, and Eirini Telali for their encouragement.
 Research at Perimeter Institute is supported by the Government of
Canada through the Department of Innovation, Science and Economic Development
and by the Province of Ontario through the Ministry of Colleges and
Universities. This work was supported by the Simons Collaboration on Celestial
Holography.

\appendix
\onecolumngrid
\input{appendices/appA_time_delay}
\input{appendices/appB_psd}
\input{appendices/appC_correlators}
\input{appendices/appD_wightman}

\twocolumngrid
\bibliography{references}
\end{document}

%% file: config/packages.tex
\usepackage{fullpage}
\usepackage{microtype} 
\usepackage[utf8]{inputenc} 
\usepackage[T1]{fontenc}
\usepackage{helvet} 
\makeatletter\def\@seccntformat#1{\protect\makebox[0pt][r]{\csname
the#1\endcsname\hspace{11pt}}}\makeatother 
\usepackage{empheq}
 
\makeatletter\renewcommand{\@dotsep}{1000} 

\usepackage{dsfont}
\usepackage{mathtools,amssymb,amsmath}
\usepackage{euscript}
\DeclareSymbolFontAlphabet{\mathbb}{AMSb}
\usepackage{bm} 
\usepackage[bb=libus, scr=boondoxo]{mathalpha}
\usepackage{braket}
\usepackage{slashed}
\usepackage{MnSymbol}

\usepackage{xcolor}
\usepackage{tcolorbox}
\usepackage{graphicx}
\usepackage{enumitem}
\usepackage{accents}
\usepackage{physics}
\usepackage{tikz}
\usetikzlibrary{decorations.pathmorphing,arrows.meta,calc}
\usepackage[a]{esvect}

\usepackage[colorlinks=true,linkcolor=blue,citecolor=purple,urlcolor = blue,linktocpage=true]{hyperref}
\numberwithin{equation}{section}

%% file: config/macros.tex
\renewcommand{\vec}[1]{{\mathbf{#1}}}
\renewcommand{\a}[2]{a_{#1,#2}}
\renewcommand{\aa}[2]{a_{#1,#2}}
\newcommand{\ad}[2]{a^{\dagger}_{#1,#2}}

\newcommand{\pn}{ \vec{p} \cdot \vec{n}}

\newcommand{\cre}[1]{a_{\vec{#1}}^\dagger}
\newcommand{\ann}[1]{a_{\vec{#1}}}

\newcommand{\Ep}[1]{|\vec{#1}|}
\newcommand{\measure}[1]{\frac{\rd^3 \vec{#1}}{(2\pi)^3 \sqrt{2\Ep{#1}}}}
\newcommand{\measuretwo}[1]{\frac{\rd^3 \vec{#1}}{(2\pi)^3 2\Ep{#1}}}
\newcommand{\dVp}[1]{\rd V_\vec{#1}}
\newcommand{\threedelta}[2]{\delta^{(3)}(\vec{#1} - \vec{#2})}

\newcommand{\Gbv}[0]{\bar{\mathcal{G}}^{(0)}_{abcd}(p)}
\newcommand{\Gv}[0]{\mathcal{G}^{(0)}_{abcd}(p)}
\newcommand{\Gvv}[0]{\mathcal{G}^{(0)}}
\newcommand{\Gs}[1]{\mathcal{G}^{(#1)}_{abcd}(p)}

\newcommand{\TTbv}[0]{\bar{\mathcal{T}}^{(0)}_{abcd}(p)}
\newcommand{\TTbvv}[0]{\bar{\mathcal{T}}^{(0)}}
\newcommand{\TTv}[0]{\mathcal{T}^{(0)}_{abcd}(p)}
\newcommand{\TTbs}[1]{\bar{\mathcal{T}}^{(#1)}_{abcd}(p)}

\newcommand{\TTbb}[0]{\bar{\mathcal{T}}^{(\beta)}_{abcd}(p)}
\newcommand{\TTb}[0]{\mathcal{T}^{(\beta)}_{abcd}(p)}

\newcommand{\TTbz}[0]{\bar{\mathcal{T}}^{(\zeta)}_{abcd}(p)}

\newcommand{\dPhi}[2]{\rd \Phi_{#1 #2}(p;k)}
\newcommand{\dPhib}[0]{\rd \Phi_{\beta}(p;k)}

\newcommand{\dPhipp}[0]{\rd \Phi_{++}(p;k)}
\newcommand{\dPhimm}[0]{\rd \Phi_{--}(p;k)}
\newcommand{\dPhipm}[0]{\rd \Phi_{+-}(p;k)}
\newcommand{\dPhimp}[0]{\rd \Phi_{-+}(p;k)}

\newcommand{\Sig}{\Sigma_n}

\newcommand{\DTI}[0]{\Delta_\zeta^{\text{TI}}}
\newcommand{\DNTI}[0]{\Delta_\zeta^{\text{NTI}}}

\newcommand{\be}{\begin{equation}}
\newcommand{\ee}{\end{equation}}
\newcommand{\bea}{\begin{eqnarray}}
\newcommand{\eea}{\end{eqnarray}}
\newcommand{\bee}{\begin{equation} \begin{aligned}}
\newcommand{\eee}{ \end{aligned} \end{equation}}
\newcommand{\lb}{\label}
\newcommand{\sss}{\scriptscriptstyle}

\newcommand{\cA}{\mathcal{A}}
\newcommand{\cB}{\mathcal{B}}
\newcommand{\cC}{\mathcal{C}}
\newcommand{\cD}{\mathcal{D}}
\newcommand{\cE}{\mathcal{E}}
\newcommand{\cJ}{\mathcal{J}}
\newcommand{\cL}{\mathcal{L}}
\newcommand{\cM}{\mathcal{M}}
\newcommand{\cN}{\mathcal{N}}
\newcommand{\cO}{\mathcal{O}}
\newcommand{\cP}{\mathcal{P}}
\newcommand{\cQ}{\mathcal{Q}}
\newcommand{\cT}{\mathcal{T}}
\newcommand{\cW}{\mathcal{W}}
\newcommand{\cX}{\mathcal{X}}
\newcommand{\cY}{\mathcal{Y}}
\newcommand{\cZ}{\mathcal{Z}}

\newcommand{\sR}{\mathscr{R}}
\newcommand{\sQ}{\mathscr{Q}}
\newcommand{\sW}{\mathscr{W}}
\newcommand{\sS}{\mathscr{S}}
\newcommand{\sK}{\mathscr{K}}
\newcommand{\sD}{\mathscr{D}}
\newcommand{\sO}{\mathscr{O}}
\newcommand{\sU}{\mathscr{U}}
\newcommand{\sZ}{\mathscr{Z}}


\newcommand{\mK}{\mathsf{K}}
\newcommand{\bmK}{\bar\mathsf{K}}
\newcommand{\mN}{\mathsf{N}}

\newcommand{\M}{\mathscr{M}} 
\newcommand{\N}{\mathscr{N}} 
\renewcommand{\H}{\mathscr{H}} 
\newcommand{\sC}{\mathscr{sC}} 

\renewcommand{\S}{\mathscr{S}} 
\newcommand{\SN}{\S_{\sU}} 

\newcommand{\Neq}{\stackrel{\sss \N}{=}} 
\newcommand{\Heq}{\stackrel{\sss \H}{=}} 
\newcommand{\EOM}{\hat{=}} 

\newcommand{\bvol}{\bm{\epsilon}} 
\newcommand{\volH}{\bm{\epsilon}_{\sss \bm{\H}}} 
\newcommand{\volN}{\bm{\epsilon}_{\sss \bm{\N}}}
\newcommand{\volM}{\bm{\epsilon}}
\newcommand{\volS}{\bm{\epsilon}_{\sss \bm{\S}}}

\newcommand{\rd}{\mathrm{d}} 
\newcommand{\Lie}{\mathcal{L}} 
\newcommand{\exd}{\bm{\rd}} 
\newcommand{\dH}{\mathbb{d}} 
 \newcommand{\LH}{\mathbb{L}} 

\renewcommand{\bar}{\overline}
\renewcommand{\a}{{\alpha}} 
\renewcommand{\b}{{\beta}} 
\newcommand{\ba}{\bar{\alpha}} 
\renewcommand{\v}{v} 
\newcommand{\ac}{\varphi} 
\newcommand{\vor}{w} 
\newcommand{\p}{\wp} 
\newcommand{\E}{\EuScript{E}} 
\newcommand{\NN}{\EuScript{N}} 
\newcommand{\A}{\EuScript{A}} 
\newcommand{\J}{\EuScript{J}} 
\renewcommand{\P}{\EuScript{P}} 
\newcommand{\T}{\EuScript{T}} 
\newcommand{\Wein}{\mathsf{N}} 
\newcommand{\EM}{\mathsf{T}} 
\newcommand{\btheta}{\bar{\theta}{}} 
\newcommand{\s}{\sigma} 
\newcommand{\bs}{\bar{\sigma}{}} 
\newcommand{\RC}{\sR} 
\newcommand{\W}{W} 
\newcommand{\xit}{\hat{\xi}}
\newcommand{\sv}{\mathrm{v}}

\newcommand{\e}{\mathrm{e}} 
\newcommand{\la}{\langle}
\newcommand{\ra}{\rangle}
\newcommand{\pa}{\partial}
\newcommand{\wh}[1]{\widehat{#1}{}}
\newcommand{\mr}{\mathring}

\newcommand{\magenta}[1]{\color{magenta}#1 \color{black}}
\newcommand{\blue}[1]{\color{blue}#1 \color{black}}
\newcommand{\red}[1]{\color{red}#1 \color{black}}
\newcommand{\PJ}[1]{\color{purple}#1\color{black}}
\newcommand{\Lau}[1]{\color{blue}#1\color{black}}

%% file: sections/introduction.tex
\section{Introduction}

 Can low-energy experiments probe fundamental aspects of quantum gravity, such as the regularization of UV divergences? 
 This question\footnote{It should not be confused with the  program which aims to establish whether the low energy gravitational field itself exhibits quantum properties through gravity-mediated entanglement experiments  \cite{Bose:2017nin, Marletto:2017kzi, Huggett:2022uui}. Our work assumes that low-energy gravity behaves quantum mechanically.} is central to discovering quantum gravity and calls for a reevaluation of how short-distance physics appears at accessible energies. Traditionally, the link between low- and high-energy physics is understood via effective field theory (EFT), which integrates out high-energy modes based on the assumption that their scales decouple. This principle successfully explains diverse phenomena in particle physics, condensed matter, cosmology, and more. Its predictive power comes from expanding observable contributions by the ratio of the observed energy to a relevant high-energy scale. This organization, however, may not work for all observables.  

Once quantum effects of gravity are included, it is no longer clear that the
scale decoupling principle holds universally. Numerous arguments
\cite{cohenEffectiveFieldTheory1999, Berglund:2022qcc,
	Peet:1998wn, Minwalla:1999px,
amelino-cameliaBroaderPerspectivePhenomenology2025}  suggest that
quantum-gravitational effects can lead to nontrivial correlations between UV and
IR scales, thereby violating the scale separation underlying standard EFT.
Concrete manifestations of this idea appear in several contexts: entropy bounds
in gravity, where the number of degrees of freedom scales with the area rather
than the volume of a region, implying a coupling between UV and IR cutoffs
\cite{tHooft:1993dmi, Susskind:1994vu, cohenEffectiveFieldTheory1999};
holographic dualities, where bulk physics is encoded in boundary  data implying
a drastic reduction of the number of bulk degrees of freedom compared to
standard EFT counting \cite{Peet:1998wn, Bousso:2002ju,Harlow:2025pvj,
Marolf:2014yga};  T-duality symmetry  in string theory  which fundamentally
connects the UV and IR \cite{Freidel:2015pka,Abel:2024twz}; and noncommutative
field theories where UV and IR divergences become interlaced
\cite{Minwalla:1999px, Grosse:2004yu}. As a concrete observable, it has been
argued in \cite{Freidel:2022ryr, Becker:2020mjl, Ferrero:2025ugd}
that a UV/IR relation rooted in a cutoff in the number of total states
accessible by the theory and connected to the gravitational entropy bound can
account for the observed vacuum energy density of the universe, a quantity that
is not natural within EFT.  Interplays between UV and IR scales also appear in
the hierarchy problem  \cite{Wells:2025hur} and are firmly established in the
physics of strongly correlated matter systems \cite{Mandal:2014cma,
Balut:2024aru}. This suggests that quantum gravitational effects may be
unavoidable even in some low-energy observables, and the challenge is to turn
these ideas into an experiment.

A variety of proposals have suggested that Michelson-type table-top laser
interferometers may be sensitive to an irreducible background quantum noise of
geometric origin \cite{hogan_interferometers_2012, Amelino-Camelia:1999vks,
	Ng:1999hm, parikhSignaturesQuantizationGravity2021,
Verlinde:2019xfb}. Several of these ideas have
motivated concrete experimental implementations, including both completed and
ongoing efforts \cite{Holometer:2016ipr,vermeulen_photon_2024, Patra:2024eke}.
From the conventional EFT perspective, however, gravitational vacuum noise is
expected to be strongly suppressed by the Planck scale, rendering such effects
observationally negligible \cite{Carney:2024wnp}. The observation of
quantum gravitational effects in this context could therefore be only explained
either by state-dependent enhancement such as squeezing
\cite{parikhSignaturesQuantizationGravity2021} or a fundamental coupling of the
UV and IR scales \cite{Verlinde:2019xfb,
amelino-cameliaBroaderPerspectivePhenomenology2025}. A key question is therefore
whether some experimentally relevant observable challenges the EFT argument and
consequently requires a drastic modification of how UV and IR scales enter the
description.  Addressing this question calls for a phenomenological framework
for quantum gravitational interferometry that identifies the appropriate
gauge-invariant geometric observable and clarifies its sensitivity to possible
failures of scale decoupling. This creates an opportunity for direct interplay
between fundamental theory and precision interferometry.

A concrete realization of a UV/IR coupling coming from entropy bound has been
proposed by Verlinde and Zurek (VZ) in \cite{Verlinde:2019xfb, Verlinde:2022hhs,
Verlinde:2019ade}. They argue that gravitational entropy acts as a regulator of
UV divergences in the modular Hamiltonian fluctuations, which are fundamentally
connected to geometric fluctuations within finite spacetime regions.  Therefore,
the coupling of the UV and IR scales does not only enter through the counting of
degrees of freedom, but also manifests directly as potentially observable
fluctuations of spacetime geometry—an insight that will guide our approach in
the following sections. 

Turning the VZ effect into an observable is a challenge that requires a phenomenological model. One such model was introduced in \cite{li_interferometer_2023, Zurek:2020ukz} as an effective description of gravitational backreaction effects on the time delay measured in interferometers. 
In this model, the UV divergences to be regulated are expected to manifest as a breakdown of the semiclassical consistency criterion: quantum fluctuations of the metric fail to remain small compared to their expectation values, or simply diverge. Whether this criterion is satisfied for interferometric observables is crucial to establish. Energy backreaction providing a mechanism connecting UV and IR physics has already appeared in \cite{Afshordi:2015iza, Afshordi:2017scc} with potential consequences for interferometer noise as well \cite{Afshordi:2019xbz}. This further emphasizes the need for a rigorous account of quantum fluctuations in the context of specific observables. 
Understanding under which condition the semiclassicality criterion of energy-fluctuation is violated is a central tenet of stochastic gravity, which extends semiclassical gravity by incorporating fluctuations of the stress-energy tensor \cite{Kuo:1993if,Hu:2008rga, Hu:2020luk, hu_induced_2004,
martin_stochastic_2000, Calzetta:1993qe,
Verdaguer:2006iy,
Perez-Nadal:2009jcz, 
Calzetta:1997rf, Cho:2021gvg, ford_quantum_2022}.

In this work, we present and analyze a gauge-invariant geometric observable: the
power spectral density of interferometric time-delay fluctuations. Our work is closely related to the work of Carney et al. \cite{Carney:2024wnp}. We focus on
linear perturbations where the metric fluctuations are decomposed into intrinsic
graviton modes and fluctuations induced by the backreaction of stress-energy. We
characterize the state-dependence of the noise spectrum associated with
intrinsic graviton fluctuations and assess their detectability. For the response
induced by backreaction, we compute the interferometric response to vacuum
stress-energy fluctuations and analyze the resulting UV behavior. 

The paper is organized as follows. Section \ref{sec:psd} develops the
general framework that connects linearized metric quantum fluctuations to the
interferometer response encoded in the power spectral density. In Section
\ref{sec:2.QF}, we outline the two types of fluctuations considered: intrinsic
and induced. Section \ref{sec:3} presents a detailed analysis of the intrinsic
fluctuations for the vacuum, thermal, and squeezed states of the graviton field\footnote{We acknowledge that the vacuum state has already been computed in \cite{Carney:2024wnp}.}.
Section \ref{sec:4} studies the noise spectrum from metric fluctuations induced
by the stress-energy tensor of a massless scalar field. In Section
\ref{sec:conclusion}, we discuss our results and the necessary next steps.
Detailed derivations are provided in the appendices.

\paragraph*{Conventions and notation}
We work in four-dimensional spacetime with mostly-plus Lorentzian metric
signature $(-+++)$. Natural units are used throughout ($\hbar = c = 1$), unless
stated otherwise. We use perturbation theory over Minkowski spacetime with
expansion parameter $\kappa_c = \sqrt{32\pi G_N}$ in the classical theory, where
$G_N$ is the gravitational constant. In the quantum theory, we replace $\kappa_c
\rightarrow \kappa := \sqrt{32\pi \ell_P^2}$ with the
Planck length $\ell_P$.  Spacetime indices are denoted by
\(a,b,c,\ldots=0,1,2,3\), while spatial indices are denoted by
\(i,j,k,\ldots=1,2,3\).  Four-dimensional spacetime vectors are denoted by
regular upright letters, e.g., $p$ with components $p^a = (p^0, \vec{p})$. Our
Fourier convention is 
\begin{align}
	h_{ab}(x) &= \int \frac{\rd^4 p}{(2\pi)^4} \tilde{h}_{ab}(p)\, e^{-ipx} \\
	\tilde{h}_{ab}(p) &= \int \rd^4 x \, h_{ab}(x)\, e^{ipx} \nonumber
\end{align}

%% file: sections/sec2_spectrum.tex
\section{Spectrum of light ray quantum fluctuations}
\label{sec:psd}

In this section, we analyze a gauge-invariant observable associated with the
time delay of light rays in an interferometer caused by perturbations of the
spacetime geometry. We derive a generic formula for the spectral
density of time delay fluctuations which we will then use to analyze specific
examples such as the effects due to canonically quantized linear perturbations
as well as the response to energy backreaction.

\subsection{Michelson interferometer model}
Michelson-type interferometers work by directing a laser beam on a beamsplitter
that splits the beam into two parts going into the two arms with mirrors at
their end. After reflection at the mirrors, the two beams recombine at the
beamsplitter and are directed to the readout system where we record the
statistics of the arriving photons. Fluctuations of the optical paths in the two
arms cause modulations in the recombined beam changing the interference pattern
and thus the photon arrival statistics. 

In this paper, we model this interferometer setup in the following way.  We
consider an interferometer with a single arm pointing in the spatial direction
denoted by the spacelike unit vector $n^a=(0,\vec{n}),\, n^2=1$. The
interferometer also picks a future-directed timelike vector $t=(1, \vec{0})$,
$t^2 = -1$ for the timeflow such that $t\cdot n = 0$. This also defines the null
frame spanned by $\ell_\pm := t \pm n$ such that $\ell_+ \cdot \ell_- = -2$. The
laser beam corresponds to a null geodesic path emitted from and arriving to the
beamsplitter, which we split into outgoing $x_+$ and incoming $x_-$ components.
Both the beamsplitter ($B$) and the mirror ($M$) are modeled with timelike
geodesics $x_B$ and $x_M$ respectively. The optical path fluctuations are
represented by the proper time delay of the arrival of the incoming light ray
for a timelike observer at the beamsplitter - formally, the beamsplitter itself.
It is clear that on Minkowski spacetime, the total elapsed proper time for the
beamsplitter between the emission and arrival of the light ray is $2L$.

\subsection{Time delay due to linear perturbations}
The following is a standard textbook computation widely used in gravitational
wave interferometry \cite{fundamentals_nodate, Maggiore:2007ulw}. It
has recently been the focus of a new analysis in the context of quantum
gravitational noise \cite{li_interferometer_2023, Carney:2024wnp,
Lee:2024oxo} which we present in a way that is suitable for the quantum analysis
later. Consider linear perturbations around Minkowski background
\begin{equation}
    g_{ab} = \eta_{ab} + \kappa_c\, h_{ab} + \cO(\kappa_c^2),
\end{equation}
where $\kappa_c:= \sqrt{32 \pi G_N}$. We are looking for
deviations in the elapsed proper time for the beamsplitter between the emission
and arrival of the light ray. Our strategy is to calculate the time delay in
temporal gauge $h_{ab}^{(t)}$ defined by $h^{(t)}_{ab}t^b = 0$, and then restore
gauge invariance by constructing the diffeomorphism that maps into the temporal
gauge. In temporal gauge, the timelike geodesics are such that their tangent
vector is $t^a$ at all order, and therefore, we only have corrections to the
null geodesics. 

The background geodesic motion of the mirror (M), beamsplitter (B), incoming and
outgoing light rays are respectively encoded in time-dependent positions
$\bar{x}^a_M(\tau),\, \bar{x}^a_B(\tau),\, \bar{x}^a_\pm(\tau)$  and given by 
\begin{subequations}
    \begin{align}
        \bar{x}_{B,M}(\tau) &= \tau\, t +  x_{B,M} \\
        \bar{x}_{+}(\tau) &= \tau\, t +  (\tau-\tau_E) n + {x}_{B} \\
        \bar{x}_{-}(\tau) &= \tau\, t -   (\tau-\tau_E - 2L) n + x_B
    \end{align}
\label{eqn:geodesics}
\end{subequations}
where $\tau_E$ is the time of emission and $x_{B,M}$ are the initial position of
the beamsplitter and mirror which are related as $x_M = x_B + L \,n$.  These
background geodesics satisfy the matching conditions
\begin{subequations}\label{matching}
    \begin{align}
        \bar{x}_B(\tau_E) &= \bar{x}_+(\tau_E) \\
        \bar{x}_M(\tau_E+L) &= \bar{x}_+(\tau_E+L)= \bar{x}_-(\tau_E+L) \\
        \bar{x}_B(\tau_E + 2L) &= \bar{x}_-(\tau_E+2L)
    \end{align}
\end{subequations}
These correspond respectively, to the intersection points at the time of
emission $\tau_E$, reflection from the mirror at time $\tau_E + L$ and arrival
back at the beamsplitter at $\tau_E + 2L$. The geodesic motion to first order is
given by $x_{\pm,B,M} = \bar{x}_{\pm,B,M} +\delta x_{\pm,B,M}$ where  $\delta
x_{\pm,B,M}$ denotes the first order geodesic deviations due to $h_{ab}$.  In
temporal gauge we have that the  geodesic motion of the beamsplitter and mirror
are undeformed so that $\delta x_{B,M}(\tau)=0$. The interferometer geometry is
illustrated in figure \ref{fig:interferometer}. 

\begin{figure}[!ht]
\centering
\begin{tikzpicture}[x=1cm,y=1cm, line cap=round, line join=round, scale = 0.4]

\tikzset{
    worldline/.style={line width=1.1pt},
    ray/.style={line width=1.0pt, decorate, decoration={snake, amplitude=0.8mm, segment length=4mm}},
    dot/.style={circle, fill=black, inner sep=1.6pt},
    tdot/.style={circle, fill=blue, inner sep=2.0pt},
    lab/.style={font=\small},
    tlab/.style={font=\small, text=blue}
}

\coordinate (Bbot) at (0,-1);
\coordinate (Btop) at (0,13);
\coordinate (Mbot) at (6,-1);
\coordinate (Mtop) at (6,13);

\coordinate (emit) at (0,0);     
\coordinate (hit)  at (6,6);     
\coordinate (ret)  at (0,12);    

\draw[worldline] (Bbot) -- (Btop);
\draw[worldline] (Mbot) -- (Mtop);

\draw[ray] (emit) -- (hit);
\draw[ray] (ret)  -- (hit);

\node[dot] at (emit) {};
\node[dot] at (hit)  {};
\node[dot] at (ret)  {};

\coordinate (emitT) at (0,0);
\coordinate (hitT)  at (6,6.35);
\coordinate (retT)  at (0,12.35);

\node[tdot] at (emit) {};
\node[tdot] at (hit)  {};
\node[tdot] at (ret)  {};

\node[lab, below=10pt] at (Bbot) {$x_B(\tau)$};
\node[lab, below=10pt] at (Mbot) {$x_M(\tau)$};

\node[lab] at ($(emit)!0.55!(hit) + (0.4,-1.0)$) {$x_{+}(\tau)$};
\node[lab] at ($(ret)!0.55!(hit)  + (0.4,1.0)$) {$x_{-}(\tau)$};

\node[tlab, left=6pt]  at (emitT) {$\tau_E$};
\node[tlab, right=6pt] at (hitT)  {$\tau_E + L + \delta\tau_+(\tau_E)$};
\node[tlab, left=6pt]  at (retT)  {$\tau_E + 2L + \delta\tau(\tau_E)$};

\coordinate (O) at (8,2.0);
\draw[-{Stealth[length=2.2mm]}, line width=0.9pt] (O) -- ++(0,1.6) node[lab, above] {$t$};
\draw[-{Stealth[length=2.2mm]}, line width=0.9pt] (O) -- ++(1.8,0) node[lab, right] {$n$};

\end{tikzpicture}
\caption{Spacetime diagram of a null ray crossing the interferometer in the $t-n$ plane. Straight black lines indicate the beamsplitter $x_B$ and the mirror $x_M$ timelike geodesics. Wiggly lines correspond to the outgoing $x_+$ as well as the incoming $x_-$ null geodesics. Blue dots and labels indicate the perturbed intersection proper times.}
\label{fig:interferometer}
\end{figure}

We are interested in evaluating the time delay $\delta \tau(\tau_E)$ for a ray
emitted at time $\tau_E$.  This time delay splits into a time delay accumulated
along the outgoing null geodesic and along the incoming null geodesics  
\begin{equation}
    \delta\tau(\tau_E)= \delta\tau_+(\tau_E)+\delta\tau_-(\tau_E)
\end{equation}
The matching conditions generalizing \eqref{matching} ensure that the
beamsplitter geodesic intersect the null geodesics. To first order, they read
\begin{subequations}
    \begin{align}
        \delta\tau_+(\tau_E)&= \ell_+\cdot \left[ (\delta x_+-\delta x_M)(\tau_E+L) \right. \\
        & \qquad \left. - (\delta x_+ - \delta x_B)(\tau_E)\right], \nonumber\\
        \delta\tau_-(\tau_E)& = \ell_-\cdot \left[ (\delta x_--\delta x_B)(\tau_E+2L) \right.\\
        & \qquad \left. -(\delta x_- -\delta x_M)(\tau_E+L)\right] \nonumber
    \end{align}
\end{subequations}
Evaluating this in the temporal gauge gives 
\begin{equation}
        \delta\tau_\pm(\tau_E) = \kappa_c \int_{\gamma_\pm}
            h_{\pm}(\tau) \rd \tau
    \label{eqn:timedelay1}
\end{equation}
where the integration regions are $\gamma_+ = [\tau_E, \tau_E + L]$ and
$\gamma_- = [\tau_E + L, \tau_E + 2L]$, while the integrand is 
\begin{equation}
    h_\pm(\tau) \equiv 
        \frac{1}{2} \ell_\pm^a \ell_\pm^b  
                h^{(t)}_{ab}(\bar{x}_\pm(\tau)) \label{eqn:hpmt}
\end{equation}
 The detailed calculation can be found in appendix
\ref{app:A}.

We know that the time delay is a gauge-invariant observable\footnote{It can be
understood as the sum of a Doppler effect due to the relative motion of the
mirror and beamsplitter, plus a Shapiro time delay along the null geodesic plus
an Einstein effect due to the beamsplitter acceleration. See \cite{Lee:2024oxo}
for more details.} which means that we should be able to express it in a gauge
invariant manner.
To do so, we construct a vector field $\zeta^a[h]$ which maps any metric $h$ to
the temporal gauge metric through the relation 
\be 
h_{ab}^{(t)}= h_{ab} -\pa_a \zeta_b[h] -\pa_b \zeta_a[h].
\ee 
The temporal gauge conditions  $h_{at}^{(t)}=0$ provides differential equations for the gauge lift vector 
\be 
h_{tt}= 2 \partial_t \zeta_t[h], \qquad
h_{ti}= \partial_i \zeta_t[h] + \partial_t \zeta_i[h],
\ee 
where $i$ is a spacelike index.  These equations can be easily solved in terms
of the operator $\pa_t^{-1}$ as 
\begin{align}
   \zeta_t(x) &= \frac{1}{2}\partial_t^{-1}h_{tt}, \qquad  
   \zeta_i(x) = \partial_t^{-1}h_{ti} - \frac{1}{2}\partial_t^{-2}\partial_{i} h_{tt}.
\end{align}
The operator $\partial_t^{-1}$ defined as 
\begin{equation}
    \partial_t^{-1} f(t,x^i) := \int_{-\infty}^t \rd t' f(t',x^i),
\end{equation}
is gauge invariant under all background diffeomorphisms that vanish at
$t=-\infty$. It can equivalently be written through its action under Fourier
transform of the function $f(x)$ as
\begin{equation}
    \int \rd^4 x\, e^{ipx} \partial_t^{-1}f(x) := \frac{i}{(p_0 + i\epsilon)} \tilde{f}(p).
    \label{eqn:ieps}
\end{equation}
This means that the operators $h_\pm$ entering the time delay integrals can be
written in a gauge invariant manner as 
\begin{equation}
    h_\pm(\tau) := 
        \frac{1}{2} \ell_\pm^a \ell_\pm^b 
            \bigg( 
                h_{ab} - \partial_a \zeta_b[h] - \partial_b \zeta_a[h] 
            \bigg)(\bar{x}_\pm(\tau)) \label{eqn:hpm}
\end{equation}
which is also related to the Riemann tensor as $\pa_t^2 h_\pm =
R_{tntn}(x_\pm(\tau))$.

Introducing the Fourier transform of $h_\pm(\tau)$ with respect to the full
spacetime coordinate\footnote{The factor  $1/p_0$ creates a IR divergence that
is regulated via an $i\epsilon$ prescription. Indeed as we just explained in
\eqref{eqn:ieps}, this factor should be understood as $1/(p_0+i\epsilon)$.}, we
write
\begin{equation}
	\tilde{h}(p) := \int_{\mathbb{R}^4} \rd^4 x\, e^{ipx} h(x) = \Sigma_n^{ab} \tilde h_{ab}
    \label{eqn:fourierhp}
\end{equation}
with the tensor
\begin{equation}
	\Sigma^{ab}_n(p) := \frac{1}{2}\sigma^a_n(p) \sigma^b_n(p)
    \label{eqn:sigma}
\end{equation}
and
\begin{equation} 
    \sigma_n^a(p) := \frac{(p\cdot n) t^a - (p\cdot t) n^a}{(p\cdot t)}.
    \label{eqn:sigman}
\end{equation}
We will refer to this tensor as the \emph{geodetic} tensor. It is of rank one
and associated with a null geodesic as seen by a pair of mirror and
beamsplitter.\footnote{We acknowledge that it was presented in a poster session
at the Aspen conference ``Observables in Quantum Gravity: From Theory to
Experiment''.}. The vector $\sigma_n^a$ satisfies 
\begin{equation}
	\sigma_n^a(p) p_a=0, \qquad 
	\sigma_n^2 = \frac{(p\cdot \ell_+)(p\cdot \ell_-)}{(p\cdot t)^2}.
\end{equation}
The first equality is the gauge invariance of the observable, which then also
implies
\begin{equation}
    \Sigma_n^{ab}(p)p_a = 0
    \label{eqn:gaugeinv}
\end{equation}
The second equality tells us that the causal genre of $\sigma_n$ is connected to
the genre of $\hat{p}^a:=  (-t^a t_b + n^a n_b) p^b$, which is the projection of
$p$ along the interferometric $(t,n)$ plane.  $\sigma_n$ is  timelike when
$\hat{p}$ is spacelike and spacelike when $\hat{p}$ is timelike. Its norm
diverges when $\hat{p}$ becomes orthogonal to $t$.\footnote{More precisely in
the limit where $p\cdot t\to0$ while $p\cdot n\neq 0.$} It only depends on
direction $n$ and not on whether we choose the outgoing or the incoming tangent
vector $\ell_\pm$. 

We also calculate the Fourier transform of the time delay
(\ref{eqn:timedelay1}), which can be rewritten using $\tilde{h}(p)$ as
\begin{align}
     \delta \tilde{\tau} (\omega) &:=\int_{-\infty}^{\infty} \rd \tau_E \,
    e^{-i\omega \tau_E} \delta {\tau}(\tau_E) \nonumber\\
    &= \kappa_c \int \frac{\rd^3 \vec{p}}{(2\pi)^3} \,
R(\omega,\vec p) \tilde{h}(p)
    \label{eqn:timedelayFkernel}
\end{align}
with $R(\omega, \vec{p})$ being the response function
\begin{align}
	&2\pi \delta(p^0-\omega) R(\omega, \vec{p}) \nonumber \\
	&:=\sum_{\alpha =
	\pm}\int_{-\infty}^\infty \rd \tau_E \, e^{-i\omega \tau_E}
																					\int_{\gamma_\pm} \rd \tau \, e^{-i \bar{x}_\pm(\tau)\cdot p} 
    \label{eqn:Fkernel}
\end{align}
where the $\tau$ and $\tau_E$ integrals together give the Dirac delta that
imposes $p^0 = \omega$. We can split this kernel into the incoming and outgoing
contributions $R = R_+ + R_-$ and using the explicit form of the background
geodesics (\ref{eqn:geodesics}), they take the form
\begin{subequations}
    \begin{align}
        R_+(\omega, \vec{p}) &= \left(
     \frac{e^{i(\omega  - \vec{p}\cdot \vec{n})L}-1}{i(\omega  - \vec{p}\cdot \vec{n})}
     \right)e^{-i \vec{p}\cdot \vec{x}_B} \\
        R_-(\omega, \vec{p}) &= \bigg( \frac{e^{2iL \omega } -e^{i(\omega - \vec{p}\cdot \vec{n})L}}{i(\omega + \vec{p}\cdot \vec{n})}\bigg) e^{-i\vec{p}\cdot \vec{x}_B}
    \end{align}
		\label{eqn:Fkernelfinal}
\end{subequations}
More details on this computation can be found in appendix \ref{app:B.1}.
The time delay in momentum space can thus be written as
\begin{equation}
        \delta \tilde{\tau}_\pm (\omega) = \kappa_c \int \frac{\rd^3 \vec{p}}{(2\pi)^3} 
     R_\pm(\omega, \vec{p})
      \, \Sigma^{ab}_n \tilde{h}_{ab}(\omega,\vec{p})
    \label{eqn:dtauomega}
\end{equation}
where $\tilde{h}_{ab}(\omega,\vec{p}) \equiv \tilde{h}_{ab}(p^a = (\omega,
\vec{p}))$. This concludes our classical analysis of the time delay to first
order. The final expression (\ref{eqn:dtauomega}) makes it explicit that the
time delay decomposes into two distinct components. On the one hand, we have the
response function $R_\pm$ that encodes the properties of the background geodesic
paths.  On the other hand, the geodetic tensor $\Sigma^{ab}_n$ governs the
contraction of the metric perturbations relevant to the observable.

\subsection{Noise power spectral density}
\label{sec:2.PSD}

In general relativity, the round--trip time of null rays is a well-defined
observable once an on-shell spacetime metric is specified.
Fluctuations of this time delay quantify how variations in the spacetime
geometry affect light propagation. In gravitational-wave interferometers, such
time-delay fluctuations are operationally accessed through variations in the
output optical power, which encodes the detector’s response to spacetime strain
\cite{Maggiore:2007ulw, fundamentals_nodate}.

The sensitivity of these measurements is ultimately limited by the ability to
resolve optical power fluctuations against a variety of noise sources. These
include environmental disturbances as well as intrinsic quantum effects in the
measurement apparatus, most notably the statistical fluctuations in photon
arrival times commonly referred to as shot noise. This noise budget
can be—and has been—accurately characterized without explicit reference to
dynamical spacetime degrees of freedom \cite{buikema_sensitivity_2020}.

In this work, we adopt a reversed perspective and ask whether—and in what
sense—the operational definition of length fluctuations is subject to
fundamental limitations arising from quantum properties of spacetime. This
question has been explored from a variety of angles in the literature, including
both theoretical proposals and past or ongoing experimental efforts 
\cite{li_interferometer_2023, vermeulen_photon_2024, hogan_interferometers_2012,
	chouInterferometricConstraintsQuantum2017,
kwonInterferometricTestsPlanckian2016a,
Richardson:2020snt,
carneyTabletopExperimentsQuantum2019, Carney:2024wnp,
parikhSignaturesQuantizationGravity2021, Sivaramakrishnan:2025srr}. 
The extent to which a genuine spacetime contribution to interferometric noise
can be defined in a way that implies an observable prediction remains an open
question.

From this perspective, the central observable of interest is the \emph{noise power
spectral density}, which characterizes fluctuations as measured in an
interferometer while remaining sensitive to the underlying spacetime dynamics. A
standard operational definition of this quantity relies on finite-time
measurements of the process, typically implemented through a window function
$w_T$ corresponding to a measurement duration $T$
\cite{papoulis2002probability}. This motivates the introduction of the windowed
Fourier transform of the time delay,
\begin{equation}
	\delta\tilde{\tau}_T(\omega) := \int_{-\infty}^\infty \rd t\, w_T(t)\,
		e^{-i\omega t} \delta \tau(t) \, 
\end{equation}
where $\omega_T(t)$ vanishes or decays rapidly outside a time interval.
One may then define the distribution of power across frequencies as measured over
a finite duration $T$ by
\begin{equation}
	S_T(\omega) := \frac{1}{\| w_T \|^2} \big\langle
	\delta\tilde{\tau}_T(\omega)\delta
	\tilde{\tau}_T^\dagger(\omega) \big\rangle ,
	\label{eqn:ST}
\end{equation}
where $\|w_T\|^2 = \int \rd t\, |w_T(t)|^2$ denotes the norm of the window
function.

The introduction of a finite measurement time adds an additional conceptual
layer to the problem and raises the question of whether such a scale could significantly affect the observable and play a fundamental role in the underlying theory. In this work, however, we follow the
customary EFT approach and focus on the infinite-time limit of the finite-duration
spectrum,
\begin{equation}
	S(\omega) := \lim_{T \to \infty} S_T(\omega) .
	\label{eqn:Sdef}
\end{equation}
For stationary random processes, for which $\langle \delta\tau(t)\delta\tau(t')
\rangle = \langle \delta\tau(t-t')\delta\tau(0)\rangle$, the Wiener--Khinchin
theorem guarantees that the autocorrelation function admits a positive spectral
measure. When this measure possesses a density, it is given precisely by the
function $S(\omega)$ defined above, which can be expressed as
\begin{equation}
	\boxed{
    S(\omega) =
    \int_{-\infty}^{\infty} \frac{\rd \omega'}{2\pi}
    \langle \delta \tilde{\tau}(\omega)\, \delta \tilde{\tau}(\omega') \rangle
	}
	\label{eqn:Stau}
\end{equation}
and which will serve as the primary working formula throughout this paper.
Equivalently, the spectrum may be written in the time domain as
\begin{equation}
    S(\omega) =
    \int_{-\infty}^{\infty} \rd \tau_E\,
    e^{-i\omega \tau_E}
    \langle \delta \tau(\tau_E)\, \delta \tau(0) \rangle .
\end{equation}
To clarify how the infinite-time limit in (\ref{eqn:Sdef}) leads to the
representation (\ref{eqn:Stau}), we provide a proof in
appendix \ref{app:B:wienerkhinchin}, together with a discussion of a particular
non-stationary case relevant for later results.

The noise spectral density provides a compact characterization of stochastic
fluctuations by describing how their variance is distributed across frequencies.
As such, it is the primary quantity used in both theoretical noise modeling and
experimental data analysis for laser interferometers.  The definition of
$S(\omega)$ relies on an expectation value $\langle \cdot \rangle$, whose
interpretation depends on the physical origin of the fluctuations under
consideration. In the context of quantum spacetime-induced noise, specifying this
expectation value is nontrivial and constitutes a key conceptual step in
determining whether such effects are observable. 

Using the recent form of the time delay (\ref{eqn:timedelayFkernel}) the
spectral density takes the generic form
\begin{align}
	S[\Delta](\omega) &= 
				\int \frac{\rd^3 \vec{p}}{(2\pi)^3}
        \int \frac{\rd^3 \vec{p}'}{(2\pi)^3}
        \int \frac{\rd \omega'}{2\pi} \\
				&\times R(\omega, \vec{p}) R(\omega', \vec{p}')\,\Delta((\omega,
				\vec{p}),(\omega', \vec{p}')) \nonumber
    \label{eqn:SGgen}
\end{align}
where $R = R_+ + R_-$ was defined in (\ref{eqn:Fkernel}), and we introduced the
fluctuation scalar $\Delta$, which is the quantum two-point function of
the gauge-invariant scalar contraction of the metric perturbation on the state
$\rho$
\begin{equation}
	\Delta(p, p') := \kappa_c^2 \Sigma^{ab}_n(p) \Sigma^{cd}_n(p')
	\langle \tilde{h}_{ab}(p) \tilde{h}_{cd}(p')\rangle_\rho
	\label{eqn:Deltapp}
\end{equation}
Translation-invariant states give rise to stationary fluctuations, and thus the
fluctuation scalar becomes
\begin{equation}
	\Delta(p,p') = (2\pi)^4\delta^{(4)}(p+p')\Delta(p)
\end{equation}
with
\begin{align}
	\Delta(p) &:= \int \frac{\rd^4 p'}{(2\pi)^4} \Delta(p,p')
	\label{eqn:fluctscalar} \\
						&= \kappa_c^2 \Sigma^{ab}_n(p) \Sigma^{cd}_n(p)
	\int \frac{\rd^4 p'}{(2\pi)^4} \langle \tilde{h}_{ab}(p)
	\tilde{h}_{cd}(p')\rangle_\rho \nonumber
\end{align}
where we used that $\Sigma^{ab}_n$ is an even function of $p$. Therefore, the
spectrum takes the generic form
\begin{equation}
    S[\Delta](\omega) = 
        \int \frac{\rd^3 \vec{p}}{(2\pi)^3}
            \,|R(\omega, \vec{p})|^2 \,\Delta(\omega, \vec{p})
    \label{eqn:SG}
\end{equation}
where we used $R(-\omega,-\vec{p}) = R^*(\omega,\vec{p})$.
Based on the split of $R$ into $R_\pm$, the spectral density correspondingly splits
into parts where we correlate the outgoing ($+$) or the incoming ($-$)
contributions to give $S = S_{++}+S_{--}+S_{+-}+S_{-+}$.  Taking the explicit
form of the response function $R$ from \eqref{eqn:Fkernelfinal}, the spectral density can be written in the following
simple form as detailed in appendix \ref{app:B.1}
\begin{widetext}
\begin{subequations}
\begin{empheq}[box=\fbox]{align}
        S_{++}[\Delta](\omega)&= 4\int_{\mathbb{R}^3} \frac{\rd^3 \vec{p}}{(2\pi)^3}
        \left(
        \frac{\sin\left((\omega  - \vec{p}\cdot \vec{n}\right)L/2)}{(\omega  - \vec{p}\cdot \vec{n}) }
         \right)^2 \Delta(\omega, \vec{p}), \\
         S_{--}[\Delta](\omega)&= 4 \int_{\mathbb{R}^3} \frac{\rd^3 \vec{p}}{(2\pi)^3}
        \left(
        \frac{\sin\left(
        (\omega  + \vec{p}\cdot \vec{n})L/2\right)}{(\omega  + \vec{p}\cdot \vec{n}
        )}
        \right)^2 \Delta(\omega, \vec{p}), \\
        S_{+-}[\Delta](\omega)
        &= 2 \int_{\mathbb{R}^3} \frac{\rd^3 \vec{p}}{(2\pi)^3} e^{-i \omega L}
        \left(
        \frac{ \cos (\vec{p}\cdot \vec{n}L) -  \cos (\omega L) }{(\omega^2  - (\vec{p}\cdot \vec{n})^2)}
        \right)
         \Delta(\omega, \vec{p}), \label{S+-}\\
        S_{-+}[\Delta](\omega)&= 2
        \int_{\mathbb{R}^3} \frac{\rd^3 \vec{p}}{(2\pi)^3}  e^{ i \omega L}
        \left(
        \frac{ \cos (\vec{p}\cdot \vec{n}L) -  \cos (\omega L) }{(\omega^2  - (\vec{p}\cdot \vec{n})^2)}
        \right)
        \Delta(\omega, \vec{p}).
    \end{empheq}
    \label{eqn:S}
\end{subequations}
\end{widetext}
where $\Delta(\omega, \vec{p}) \equiv \Delta\big(p^a=(\omega, \vec{p})\big)$.
Note that the off-diagonal terms $S_{\pm \mp}$ are complex conjugates of each
other, and due to spatial translation-invariance $\Delta(\omega, -\vec{p}) =
\Delta(\omega, \vec{p})$, the diagonal contributions are identical $S_{++} =
S_{--}$. The result \eqref{eqn:S} holds for any function $\Delta$ that comes
from a translation-invariant state $\rho$. For a state that is not
translation-invariant, such as a squeezed state, one has to go back to the definition of the spectral
density (\ref{eqn:Sdef}) and compute the spectrum accordingly.
The squeezed state example will be studied in section \ref{sec:3.squeezed}. 
The choice of $\Delta$ amounts to a choice of state for the gravitational and
matter fields, and requires physical insight.
Note that $\omega$ is not the frequency of the optical field, it is the inverse timescale of the particle-flux modulation and thus sets the timescale of the interferometer response. 
The geometric optics approximation for the laser field is therefore justified, since the typical frequency of the optical field is much larger than the probed fluctuation scale $\omega$ and smaller than the Planck scale.

\subsection{Quantum fluctuations} \label{sec:2.QF}
We now set up our framework for quantum fluctuations that we will use to compute
the fluctuation scalar $\Delta$ for different states. Starting with the Einstein
equations $G_{ab}=8\pi G_N T_{ab}$, we use that when expanded to first order in
$G_N$, they can be written in harmonic gauge $\pa_b h^{ba}=0$ as
\cite{blanchet_post-newtonian_2024, poisson2007post}
\be \label{eqn:Eequation}
 \Box h_{ab}= -\frac{\kappa_c}{2} \bar{T}_{ab}
\ee 
where $\Box=\eta^{ab}\pa_a \pa_b$ is the background d'Alembert operator,
$T_{ab}$ is a stress tensor satisfying the conservation law $\pa_b T^{ab}=0$ and
$\bar{T}_{ab}= T_{ab}-\frac12 \eta_{ab} T$ is the trace-reversed stress tensor.
$T_{ab}$ is the total stress tensor  which is the sum of the matter stress
tensor and a gravitational one $T_{ab}= T_{ab}^{\text{matter}} +
T_{ab}^{\text{grav}}$, where $T_{ab}^{\text{grav}}$ is quadratic in $h$.

The perturbative quantization of gravity means that we treat the equation
\eqref{eqn:Eequation} as an operatorial equation.  At the order that we are
working with, this means that  we can  decompose the quantum metric operator in
the Heisenberg picture as a sum of a homogeneous solution and a retarded
sourced solution $\hat{h}_{ab}= \hat{h}^{\text{H}}_{ab} + \kappa_c \hat{H}_{ab}$
such that 
\begin{subequations}
\begin{align}
	&\Box \hat{h}^H_{ab}=0  \\
	&\hat{H}_{ab}(x) = \frac12 \int \rd^3y \,  G_R(x,y) \hat{\bar{T}}_{ab}(y)
	\label{eqn:homret}
\end{align}
\end{subequations}
where $G_R$ is the retarded propagator solving 
\begin{equation}
	\Box G_R(x-y) = - \delta^{(4)}(x-y)
\end{equation}
The total metric fluctuation depends on a choice of state
$\rho$ in the total Hilbert space $ \mathcal{H}_{\text{GR}}\otimes
\mathcal{H}_{\text{mat}}$, where $\mathcal{H}_{\text{GR}}$ is the Hilbert space
of linearized gravitational fluctuation and $\mathcal{H}_{\text{mat}}$ the
Hilbert space of linearized matter fluctuations.  This implies that the total
metric fluctuation is, at this order, the sum of an \emph{intrinsic fluctuation} and an
\emph{induced fluctuation} associated to $\hat{h}^H$ and $\hat{H}$ respectively. The
total fluctuation scalar $\Delta(p)$ therefore depends on this choice of state.
In this paper, we restrict our choice to states that do not change the
background, which means that $\langle \hat{h}_{ab}\rangle_\rho=0$, and that do
not mix intrinsic and induced metric perturbations, i.e. $\langle \hat{h}^H_{ab}
\hat{H}_{cd}\rangle_\rho=0$.  This means that the total fluctuation for these
states is therefore given by 
\be 
\langle \hat{h}_{ab} \hat{h}_{cd} \rangle
= \langle \hat{h}^{H}_{ab} \hat{h}^H_{cd} \rangle + \kappa_c^2 
\langle \hat{H}_{ab} \hat{H}_{cd} \rangle.
\ee
In the following we will only work with quantum operators and thus reset the
conventions and denote from now on the rescaled quantum operators simply $h
\equiv \hat{h}^{H}/\sqrt{\hbar}$ and $H\equiv \hat{H}/\sqrt{\hbar}$. This has
the effect of rescaling $\kappa_c \to \kappa =\sqrt{32\pi G \hbar} = \sqrt{32\pi \ell_P^2}$, which is
proportional to the Planck length.\footnote{A simpler way is to promote all the
classical formula and work in the unit $\hbar=1$ as is customary.} The
normalization of the field is such that the Fierz-Pauli action obtained
from the perturbative expansion of the Einstein-Hilbert action $\int \sqrt{-g}
\, R$
is given by 
\begin{equation}
    S_\text{FP}[h] =-\frac{\hbar}2  \int \rd^4 x \, \partial_a \bar{h}_{bc}\,\partial^a {\bar{h}}^{bc}.
\end{equation}
where $\bar{h}$ denotes the trace-reversed metric. Moreover, the total fluctuation can be written as 
\be 
\langle {h}_{ab} {h}_{cd} \rangle
= \langle {h}^{H}_{ab} {h}^H_{cd} \rangle + \kappa^2 
\langle {H}_{ab} {H}_{cd} \rangle.
\ee
This means that the total fluctuation scalar splits into an intrinsic and an
induced fluctuation piece 
$\Delta^H_\rho$. 
\begin{equation}
	\Delta_\rho = \Delta^h_\rho + \Delta^H_\rho
\end{equation}
The fact that the stress tensor admits a decomposition into a
matter component and a gravitational component means that the induced
fluctuation tensor further splits into a matter contribution $\Delta^{H
\text{mat}}_\rho$ and a gravitational contribution $\Delta^{H \text{gr}}_\rho$.

This concludes our discussion of the general observable. In the following we focus on the construction of both intrinsic and matter induced
fluctuation for specific states. The study of the induced gravitational contribution is postponed to future work.

%% file: sections/sec3_intrinsic.tex
\section{Intrinsic fluctuations}
\label{sec:3}

\newcommand{\Dhv}{\Delta^h_{|0\rangle}}
\newcommand{\Dhb}{\Delta^h_{\beta}}
\newcommand{\Dhz}{\Delta^h_{\zeta}}

\newcommand{\polproj}[1]{\gamma}

In this section, we analyze the spectrum of quantized intrinsic fluctuations
$h^H_{ab}$ from \eqref{eqn:homret} for the vacuum, thermal, and squeezed states.
The vacuum state spectrum was previously computed by Carney et al. in
\cite{Carney:2024wnp}. We rederive their results for completeness, and
then analyze the thermal and squeezed states which exhibit interesting effects.

The trace-reversed metric perturbation $\bar{h}^H$ solves the
homogeneous wave equation on Minkowski space 
\begin{equation}
    \Box \bar{h}^H_{ab} = 0
\end{equation}
with the harmonic gauge condition $\partial^a \bar{h}^H_{ab}=0$, and can thus be
quantized with the canonical procedure. The mode expansion is
\begin{align}
    \bar{h}_{ab}^H(x) 
		= &\sumint_{s}\measure{p}  
				\bigg(
            \epsilon_{ab}^{(s)}(p)\, \aa{p}{s} \, e^{-ipx} + 
						\text{h.c.}
        \bigg)\bigg|_{p^0=|\vec{p}|} \nonumber
\end{align}
where the creation and annihilation operators satisfy
\begin{equation}
    [\aa{p}{s}, \ad{p'}{s'}] = (2\pi)^3 \delta_{s,s'}\, \delta^{(3)}(\vec{p}-\vec{p}')
\end{equation}
The polarization tensor $\epsilon_{ab}^{(s)}(p)$ is defined by a
complete gauge fixing procedure and the normalization condition
$\epsilon^{(s)}_{ab}\epsilon^{(s')*\,ab} = \delta_{s,s'}$. The harmonic gauge
condition implies $\epsilon^{(s)}_{ab}(p)p^b = 0$, and the residual gauge
can be fixed covariantly by choosing a null vector $q$ dual to $p$:
$q\cdot p = -1$ such that $\epsilon^{(s)}_{ab}(p)q^b = 0$ as well as the
traceless condition $\eta^{ab} \epsilon^{(s)}_{ab}(p) = 0$. In temporal gauge $\epsilon^{(s)}_{ab}t^b = 0$, $q$ is in the $p-t$ plane, and it takes the form $q^a = \frac{-1}{p\cdot t}(\frac{1}{2p\cdot t}p^a + t^a)$\footnote{For on-shell $p^a=(\omega, \vec{p})$, this is $q^a=\frac{1}{2\omega}(1, -\frac{1}{\omega}\vec{p})$}. 
This uniquely defines a 2-dimensional polarization space with the corresponding projector
\begin{equation}
    \polproj{p}_{ab} := \eta_{ab} + p_a q_b + q_a p_b
    \label{eqn:polproj}
\end{equation}
These imply that the polarization sum
\begin{equation}
    \Pi_{abcd}(p):= \sum_s \epsilon^{(s)}_{ab}(p) \epsilon^{(s)*}_{cd}(p)
		\label{eqn:polsum}
\end{equation}
can be written in terms of $\gamma_{ab}$ in the following way
\begin{equation}
    \Pi_{abcd}(p) = \frac{1}{2} 
    \bigg(
        \polproj{p}_{ad} \polproj{p}_{bc} + 
        \polproj{p}_{ac} \polproj{p}_{bd} - 
        \polproj{p}_{ab} \polproj{p}_{cd}  
    \bigg)
\end{equation}
We will also need the following polarization sum $\sum_s
\epsilon^{(s)}_{ab}(p) \epsilon^{(s)}_{cd}(-p)$, which does not have a
basis-independent form as it is not preserved under $U(2)$ transformations in
the polarization index.  For this, we choose a helicity basis adapted to our
interferometer plane
\begin{equation}
    \epsilon^{(\pm)}_{ab}(\vec{p}) = e^\pm_a(\vec{p}) e^\pm_b(\vec{p})
\end{equation}
with $ e^\pm_a(\vec{p}) = \frac{1}{\sqrt{2}}(e^1_a(\vec{p}) \pm i
e^2_a(\vec{p}))$. Since $t^a=(1, \vec{0})$, we only have spatial components. The
basis vectors satisfying our conditions and constructed from $\vec{n}$ are 
\begin{equation}
    \vec{e}^1(\vec{p}) = \frac{\vec{p} \times \vec{n}}{|\vec{p} \times \vec{n}|}, \quad
    \vec{e}^2(\vec{p}) = \frac{ \vec{p}\times (\vec{p} \times \vec{n}) }{|\vec{p}\times (\vec{p} \times \vec{n})|}
\end{equation}
These satisfy $|\vec{e}^1| = |\vec{e}^2| = 1, \, \vec{e}^1 \cdot \vec{e}^2 = 0,
\,\vec{p}\cdot \vec{e}^{1,2} = 0\, \vec{q}\cdot \vec{e}^{1,2} = 0$. In this
basis, we can easily see that the complex conjugation flips the
$\vec{p}$ orientation $\epsilon^{(\pm)}_{ab}(-\vec{p}) =
\epsilon^{*(\pm)}_{ab}(\vec{p})$ and therefore
\begin{equation}
	\sum_s \epsilon^{(s)}_{ab}(p) \epsilon^{(s)}_{cd}(-p) = \Pi_{abcd}(p).
		\label{eqn:polconj}
\end{equation}

\subsection{Vacuum state}
\begin{figure}[h]
    \centering
    \includegraphics[width=1.0\linewidth]{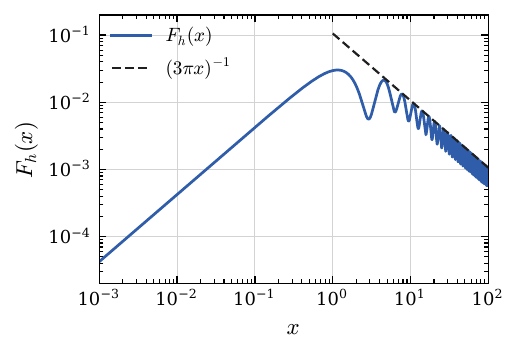}
		\caption{Dimensionless function $F_h(x)$ determining the intrinsic vacuum
		spectrum $S[\Dhv](\omega)=\kappa^2 L\,\theta(x)F_h(x)$, with $x=\omega L$. The
		dashed curve shows the large-$x$ upper envelope $(3\pi x)^{-1}$.}
    \label{fig:SDh}
\end{figure}

As a first case, we consider the spectrum for the vacuum state $\aa{p}{s}\ket{0} = 0$.
We need to calculate the fluctuation scalar $\Delta(p)$ as defined in
(\ref{eqn:fluctscalar}) which we denote by $\Dhv$ for the vacuum case and is
thus defined as
\begin{equation}
    \Dhv(p) := \kappa^2\int \frac{\rd^4 p'}{(2\pi)^4} \Sigma^{ab}_n
		\Sigma^{cd}_n \bra{0}\tilde{h}^H_{ab}(p) \tilde{h}^H_{cd}(p') \ket{0}
\end{equation}
We only need the vacuum Wightman function, which is
\begin{equation}
    W^{(0)}_{abcd}(x-x') = \int \frac{\rd^4 p}{(2\pi)^3} \delta_+(p^2)\Pi_{abcd}(p)e^{-ip(x-x')}
\end{equation}
where $\delta_+(p^2) \equiv \theta(p^0)\delta(p^2)$, and we used that $h^H_{ab} = \bar{h}^H_{ab}$ due to the traceless condition. Its Fourier transform is
\begin{equation}
    \tilde{W}^{(0)}_{abcd}(p) = 2\pi\, \Pi_{abcd}(p)\delta_+(p^2)
    \label{eqn:2pt}
\end{equation}
and therefore, the fluctuation scalar has the form
\begin{equation}
    \Dhv(p) 
    = 2\pi \kappa^2 \delta_+(p^2)\Sigma^{ab}_n\Sigma^{cd}_n \Pi_{abcd}(p)
    \label{eqn:Dh}
\end{equation}
Given the gauge invariance $\Sigma^{ab}_n
p_b = 0$, it is only the $\eta$ background metric contributions in $\Pi_{abcd}$
that survive in the contraction, and thus we get
\begin{equation}
    \Dhv(p) = \pi\kappa^2\delta_{+}(p^2) \Sigma_n^2(p)
    \label{eqn:Dhv}
\end{equation}
Furthermore, using (\ref{eqn:sigma}), the norm of the geodetic tensor is
\begin{equation}
    \Sigma_n^2 = \frac{1}{4}\bigg( 1-\frac{(p\cdot n)^2}{(p\cdot t)^2}\bigg)^2 
    \label{eqn:sigma2}
\end{equation}
We can now compute the noise spectral density by plugging $\Dhv$ into the
expressions $S[\Delta]$ from (\ref{eqn:S}), which identifies the frequency as $\omega = p^0 = -p\cdot t$ and we are left with a 3-dimensional momentum integral for each branch. Looking at the formula
(\ref{eqn:Dhv}) above, we can see that the $\delta(p^2)$ factor restricts the
$\int \rd^3 \vec{p}$ integral to an integral on a
2-sphere in momentum space where $|\vec{p}| = \omega$, and we are only left with
a 2-dimensional angular integral. It can be straightforwardly evaluated, the details of which can be found in appendix
\ref{app:B.2}. The spectrum we get is
\begin{align}
	S[\Dhv](\omega) &= \frac{\theta(\omega)\kappa^2}{4\omega\, \pi} \bigg[1 -
	\frac{3}{(\omega L)^2} + \frac{2\sin(2\omega L)}{(\omega L)^3} \nonumber\\
&\quad - \cos(2\omega L)\bigg(\frac{1}{3} + \frac{1}{(\omega L)^2} \bigg) \bigg]
  \label{eqn:SDh}
\end{align}
This is in agreement with formula (6) from \cite{Carney:2024wnp} up to a factor of $2$ and the restriction to positive frequencies.
Let us separate the dimensional part of the spectrum by introducing the positive dimensionless parameter $x :=
\omega L$ so that
\begin{equation}
    S[\Dhv](\omega) = L \,\kappa^2  \theta(x)F_h(x)
    \label{eqn:SDh2}
\end{equation}
where
\begin{equation}
    F_h(x) = \frac{1}{4\pi x}\bigg[1 - \frac{3}{x^2} + \frac{2\sin(2x)}{x^3}- \cos(2x)\bigg(\frac{1}{3} + \frac{1}{x^2} \bigg) \bigg]
		\label{eqn:Fh}
\end{equation}
A plot of this function can be found in figure \ref{fig:SDh}.
From the closed-form expression and the plot, several characteristic features of
the spectrum can be identified:
\paragraph*{Behavior in the limits.} Both the $x \rightarrow 0$ and the $x \rightarrow \infty$ limits are finite and tend to zero. For low $x$, the spectrum behaves as
\begin{equation}
    F_h(x) =  \frac{2x}{15\pi} + O(x^3)
\end{equation}
while for large $x$, the spectrum decays as $x^{-1}$, with oscillatory
modulations of the same order. More precisely, the tail of the spectrum is
asymptotically oscillating with an inversely decreasing envelope as
\begin{align}
 F_h(x)
&= \frac{1}{6\pi x}\left(1 + \sin^2 x\right) +O\!\left(\frac{1}{x^3}\right) \\
F_h(x) &\lesssim \frac{1}{3x \pi} 
\end{align}
with oscillations coming from the term $\cos(2x)/(12x\pi)$.

\paragraph*{Large-$x$ scaling and total power.} The $1/x$ decay is commonly referred to as ``pink noise'' or  $1/f$ noise \cite{mandelbrot_multifractals_1999, sornette_critical_2006}  and has the property that equal power is contained in each frequency octave, that is, $\int_\omega^{2\omega} \rd \omega'\, S_{(1/f)}(\omega')$ is independent of $\omega$ where $S_{1/f}(\omega) \sim \omega^{-1}$. In our case, this property is broken by the oscillatory behavior but is restored asymptotically in the large $x$ limit, where we can write
\begin{equation}
    \lim_{x \rightarrow \infty}\int_x^{2x} \rd x'\, F_h(x') = \frac{\ln 2}{4\pi}
\end{equation}
It is worth noting that the famous characteristic of the $1/f$ noise, the UV
divergence of the total power,  also holds here\footnote{This UV logarithmic
	divergence is discussed in \cite{Carney:2024wnp} and taken care of by replacing $S(\omega)\to S(\omega) e^{-\epsilon\omega}$. This $\epsilon$ prescription can be related to an $i\epsilon$ prescription in time $t\to t+i\epsilon$ that enters the definition of the Wightman two-point function.
}
\begin{equation}
\int_{0}^\Lambda \rd \omega\, S[\Dhv](\omega) \sim \kappa^2 \ln(\Lambda L)
    \label{eqn:powerSh}
\end{equation}
We also note that a power-law scaling is indicative of long-range temporal correlations in the underlying dynamics \cite{sornette_critical_2006}.

 \paragraph*{Low-$x$ peak.} The spectrum exhibits a global maximum at
$x\sim 1.15$. Translating this back into the dimensionful spectrum, the
position of the peak scales inversely with the arm length,
$\omega_\text{peak} \sim L^{-1}$, while its amplitude grows linearly with
$L$.

 \paragraph*{Oscillatory structure.} Beyond the overall power-law falloff,
the spectrum displays a rich oscillatory pattern coming from the
trigonometric functions.
These  may be interpreted as the appearance of
gravitationally induced constructive and destructive interference. 
Separation of peaks is well approximated by $\Delta x = n\pi,\, n =
	1,2,...$ and the corresponding frequency is the integer multiple of the
light crossing time.
    
\paragraph*{Magnitude of the noise.} 
The function $F_h$ has its maximum at $x_\text{max} \sim 1.15$, and 
the spectrum at the maximum is $S[\Dhv](\omega_\text{max})=
\kappa^2 L F_h(x_\text{max})$. This means that to maximize detection one has to
look at frequencies of order the inverse interferometer light-crossing time.

The relevant range of $x$ can be determined as
\begin{equation}
    x = 2\pi f L \sim 2\cdot10^{-8} \, \Bigg( \frac{f}{1\,\text{Hz}} \bigg) \Bigg( \frac{L}{1\,\text{m}} \bigg)
\end{equation}
For LIGO \cite{buikema_sensitivity_2020} with $L = 4\,\text{km}$ and a frequency
range of $100\, \text{Hz}$ to $5 \text{kHz}$, this gives $x$ between $0.008$ and
$0.4$. For the GQuEST \cite{vermeulen_photon_2024} table-top interferometer with
$L = 5\, \text{m}$ and a frequency range of $10-20 \, \text{MHz}$, this gives
$x$ between $1$ and $2$, showing that it is better designed to probe the peak
frequency. At the point where the signal has its maximum we have
\begin{align}
    S[\Dhv](\omega_\text{max}) &= 32\pi \ell_P^2 L F_h(x_\text{max}) \\
     &\sim \bigg(1.6\cdot 10^{-39}
     \frac{\text{m}}{\sqrt{\text{Hz}}}
 \bigg)^2 \bigg( \frac{L}{1\,
 \text{m}}\bigg) \nonumber
\end{align}
where $F_h(x_\text{max}) \sim 0.03$ and $x_\text{max}$ corresponds to
\begin{equation}
    f_\text{max} = \frac{x_\text{max}}{2\pi L}\sim 55\, \text{MHz} \, \bigg(\frac{1 \,\text{m}}{L} \bigg)
\end{equation}

\subsection{Thermal state}

Let us now analyze what happens if instead of the vacuum state $\ket{0}$, we
use a thermal state for the gravitational field. The density matrix 
\begin{equation}
    \rho_\beta = \frac{e^{-\beta H}}{Z}   
\end{equation}
with $Z= \Tr e^{-\beta H}$ and the Hamiltonian
\begin{equation}
    H = \int \frac{\rd^3 \vec{p}}{(2\pi)^3} E_p\, \sum_s \ad{p}{s}\aa{p}{s}
\end{equation}
The thermal Wightman function is a standard textbook exercise, see for example
\cite{laine_basics_2016}, it takes the form 
\begin{align}
    W^{(\beta)}_{abcd}(x) 
				&:= \Tr \bigg( \rho_\beta\, h^H_{ab}(x) h^H_{cd}(0)\bigg)
				\label{eqn:thermalWightman} \\
				&= W^{(0)}_{abcd}(x) + \delta W^{(\beta)}_{abcd}(x)  \nonumber
\end{align}
with the thermal contribution
\begin{equation}
    \delta W^{(\beta)}_{abcd}(x) 
        = \int \frac{\rd^4 p}{(2\pi)^3} \delta(p^2)
				\Pi_{abcd}(p)\,e^{-ipx}n_\beta(|p \cdot t|) 
\end{equation}
which admits contributions from both positive and negative frequencies. The Bose-Einstein distribution function
\begin{equation}
    n_\beta(E) := \frac{1}{e^{ \beta E} - 1}
\end{equation}
For the sake of completeness, we present this derivation in appendix
\ref{app:D:thermal}. The fluctuation scalar $\Delta_\beta^h$ therefore  splits into the
vacuum piece $\Dhv$ from (\ref{eqn:Dhv}) and a thermal contribution 
\begin{align}
   \Dhb(p) 
    &:= \kappa^2 \int \frac{\rd^4 p'}{(2\pi)^4} 
        \Sigma^{ab}_n\Sigma^{cd}_n \langle \tilde{h}^H_{ab}(p)\tilde{h}^H_{cd}(p') \rangle_\beta \nonumber\\
    & = \pi \kappa^2 \Sigma_n^2 \bigg( \delta_+(p^2) +
		\delta(p^2)n_\beta(|p\cdot t|) \bigg) \nonumber \\
		&= \Dhv(p)  + n_\beta(|p\cdot t|)\bigg( \Dhv(p) + \Dhv(-p) \bigg) 
    \label{eqn:Dhb}
\end{align}
where we used $\delta(p^2) = \delta_+(p^2) + \delta_-(p^2)$. The
spectrum takes the form
\begin{align}
    S[\Dhb](\omega) &= S[\Dhv](\omega) \\
										&+ n_\beta(|\omega|) \bigg( S[\Dhv](\omega) +
			S[\Dhv](-\omega) \bigg) \nonumber \\
		&= L \kappa^2  \bigg( \theta(x) F_h(x) + \, n_\beta(|x|/L) F_h(|x|)
			\bigg)
\end{align}
We can see that the same function $F_h$ appears as in the vacuum case
(\ref{eqn:Fh}), and the thermality manifests as a multiplication by a
frequency-dependent Bose-Einstein factor.  The Bose-Einstein factor depends only
on the dimensionless ratio $\beta/L$:
\begin{equation}
    n_\beta(|\omega|) =  n_\beta(|x|/L)
    = \bigg[ \exp\bigg( \frac{\beta}{L}|x|\bigg) -1 \bigg]^{-1}
\end{equation}
which takes values
\begin{equation}
    \frac{\beta}{L} \sim 2.28\cdot 10^{-3}
    \bigg( \frac{1\, \text{K}}{T} \bigg)
    \bigg( \frac{1\, \text{m}}{L} \bigg)
\end{equation}
Furthermore, as the thermal Wightman
function has both positive and negative frequency components, the support of the
spectrum is no longer restricted to positive frequencies.
The positive-frequency spectrum is related to the absorption of energy and the
negative frequency part to the emission of energy \cite{Clerk:2008tlb}. Since
the vacuum does not emit energy, only the positive--frequency
contribution appears in the vacuum case. 

\begin{figure}[h]
    \centering
    \includegraphics[width=1.0\linewidth]{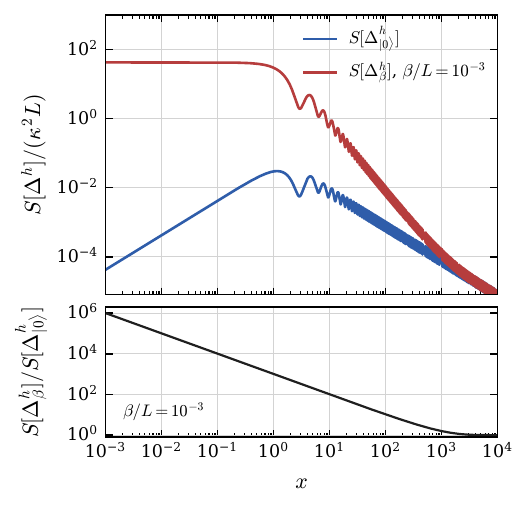}
		\caption{Comparison of the intrinsic spectra for the
		vacuum and a thermal graviton state. The upper panel shows
		$S[\Dhv]/(\kappa^2 L)=F_h(x)$ and
		$S[\Dhb]/(\kappa^2L)=F_h(x)\big[1+n_\beta(x/L)\big]$ as functions of $x=\omega
		L$, for $\beta/L=10^{-3}$. The lower panel shows the enhancement
		$S[\Dhb]/S[\Dhv]$.}
    \label{fig:SD0vsbeta}
\end{figure}

In the limit $\lim_{\beta
\rightarrow \infty} n_\beta = 0$, we recover the vacuum spectrum (\ref{eqn:SDh}). For non-zero temperature, the new term with $n_\beta$
provides an amplification of the signal, especially for small argument, as
$n_\beta(x/L) \sim L/(\beta x)$ for small $\beta x/L$. While $n_\beta( x/L )$
diverges for $x\rightarrow 0$, the factor $F_h(x)n_\beta(x/L)$ remains finite
with the limit
\begin{equation}
    \lim_{x \rightarrow 0} F_h(x)n_\beta(x/L)= \frac{2L}{15\pi \beta}
\end{equation}
which is also a good approximation for the maximum of the signal for high
temperatures. For a comparison of the spectrum to the vacuum case, see figure
\ref{fig:SD0vsbeta}.

We see that  graviton thermal states provide an amplification of the spectral density as
the inverse temperature $\beta$ decreases, that is, the hotter the heat bath the
stronger the noise. 

\noindent \paragraph*{Planckian enhancement from thermality}

Let us recall that in the canonical approach to quantum gravity, one can
decompose the gravitational perturbation into a scalar (or spin-$0$) mode, a
longitudinal (or spin-$1$) mode, and radiative (or spin-$2$) mode
\cite{Arnowitt:1962hi,Chowdhury:2021nxw,Ciambelli:2023mir}. The spin-$2$ mode
describes the linear perturbation, while the spin-$0$ and spin-$1$ modes encode
the second-order backreaction through the  Hamiltonian and diffeomorphism
constraints. In other words, the spin-$0$ and spin-$1$ components encode
fluctuations of the background geometry, which differs from the spin-$2$ and
matter EFT fluctuation. The linear perturbation we are studying in this section
are the spin-$2$ fluctuations, while the pixellon model  introduced in
\cite{li_interferometer_2023, Zurek:2020ukz} models the
spin-$0$ fluctuation. 

The pixellon model proposes that backreaction effects can be effectively
described in terms of an effective thermalization of the two-point function for
the spin-$0$ modes.  The enhancement mechanism in this effective description
comes through a high occupation number $N\sim A/\ell_{P}^2$ of low-energy
spin-$0$ modes, where $A\sim L^2$ is the area of the sphere enclosing the
interferometer. These modes have thermal energy $\beta \omega \sim 1/\sqrt{N}$
leading to the Brownian  enhancement $n_\beta \sim \sqrt{N} \sim
\frac{L}{\ell_P}$. This modifies the scaling of the noise spectrum from $S\sim
\ell_P^2 L$ to $S\sim \ell_P L^2$. For an interferometer with $L= 1\,
\mathrm{m}$, this corresponds to a noise level $S\sim (10^{-22} \,
\frac{\text{m}}{\sqrt{\text{Hz}}})^2 $, placing the effect within experimental
reach.

In our framework, the energy $E$ appearing in the occupation number $n_\beta(E)$ is naturally identified with the frequency $\omega$ probed by the interferometer, and so the only free parameter is the inverse temperature $\beta$. To reproduce the same enhancement, we can define an effective  temperature $T_\text{pix}$ by matching the high-temperature limit ($\beta_\text{pix} \to 0$) of the Bose--Einstein distribution
\begin{equation}
    n_{\beta_\text{pix}}(\omega) \to \frac{k_B T_\text{pix}}{\hbar \omega} := \frac{L}{\ell_P}.
\end{equation}
This implies 
\begin{equation}
    T_\text{pix} = T_P\,\omega L
\end{equation}
with the Planck temperature $T_P  \sim 10^{32}\,\text{K}$. In other words, the
strength of signal amplitude proposed in the (spin-$0$) pixellon model is
formally equivalent to a thermal bath of (spin-$2$) gravitons at Planckian
temperature. It is therefore unrealistic for a model involving only  thermalized
gravitons to account for a signal of such magnitude.

\subsection{Squeezed state}
\label{sec:3.squeezed}
Let us now look at the spectrum of a squeezed state. We define the squeezing
operator parametrized by the complex squeeze parameter $\zeta(p)$
\begin{equation}
    S(\zeta) := \exp\bigg( \frac{1}{2} \sumint_s \frac{\rd^3 \vec{p}}{(2\pi)^3}
    \big[ 
			\zeta^{*}(\vec{p}) \aa{p}{s}\aa{-p}{s} - \text{h.c.}
    \big]\bigg)
    \label{eqn:squeezeop}
\end{equation}
Its action on the vacuum defines the  squeezed state 
$\ket{\zeta} := S(\zeta)\ket{0}$. Note that we chose the squeezing to be
polarization-independent, and it is only the even part of $\zeta$ that
contributes $\zeta(-\vec{p}) = \zeta(\vec{p})$ since $\aa{p}{s}\aa{-p}{s} =
\aa{-p}{s}\aa{p}{s}$. It is convenient to write $\zeta = r e^{i\theta}$.  The
expectation of $h_{ab}$ in a squeezed state vanishes\footnote{This is why we
restrict our analysis to squeezed states and do not include coherent states.}
and we can focus on the two point function.

The Wightman function in a squeezed state can be easily evaluated, its
computation is detailed in appendix
\ref{app:D:squeezed}, and see also \cite{Hsiang:2024qou}
\begin{align}
		&W^{(\zeta)}_{abcd}(x,x') 
		= \int \frac{\rd^4 p}{(2\pi)^3} \delta_+(p^2) \Pi_{abcd}(p) \\
		&\quad \times \bigg[M(p) e^{ip^0(t+t')}e^{-i \vec{p}(\vec{x} - \vec{x}')} +
			\text{h.c.}\nonumber \\
		&\qquad + (1+N(p))e^{-ip(x-x')}+ N(p) e^{ip(x-x')} \bigg] \nonumber
			\label{eqn:squeezedWightman}
\end{align}
where we defined the quantities $N(p) := \sinh^2 r(p),\, M(p) :=
-\frac{e^{i\theta(p)}}{2}\sinh 2r(p)$. 
This Wightman function is not translation-invariant as the squeeze operator does
not commute with the Hamiltonian. Accordingly, 
the Fourier transform $\tilde{W}^{(\zeta)}$ evaluates 
into a sum of a time-translation-invariant contribution (TI) and a non-time-translation-invariant contribution (NTI).
The TI part of the Wightman function takes a form analogous to the thermal case 
\begin{align}
	&\tilde{W}_{abcd}^{(\zeta)\text{TI}}(p,p') 
	= (2\pi)^5\delta^{(4)}(p+p') \\
	&\quad \times \Pi_{abcd}(p)\bigg(
	\delta_+(p^2) + \delta(p^2) N(p)  \bigg) \nonumber
\end{align}
while the NTI contribution reads 
\begin{align}
	&\tilde{W}_{abcd}^{(\zeta)\text{NTI}}(p,p') 
	= (2\pi)^5 
	\delta(p'^0 - p^0) \delta^{(3)}(\vec{p} + \vec{p}') \nonumber  \\
	&\quad \times
	\Pi_{abcd}(p) 
	\bigg(\delta_+(p^2) M(p) + \delta_-(p^2) M^*(p)  \bigg) \label{eqn:WNTI} 
\end{align}
 Since stationarity was a crucial assumption in
writing the spectral density as the Fourier transform of the time delay
autocorrelation function (see the discussion at the beginning of Section \ref{sec:2.PSD}), we have to go back to the
first definition of the spectrum (\ref{eqn:Sdef}) that involved a long-time
average of a finite-time measurement. 
While $W^{\text{TI}}(t,t') = W^{\text{TI}}(t-t')$ the component $W^{\text{NTI}}$ depends on  $t+t'$.
In appendix
\ref{app:B:wienerkhinchin}, we
show that the NTI contribution to the Wightman function goes to zero in
the infinite-time limit, i.e. 
$ \lim_{T \to \infty} S_T^{\text{NTI}}(\omega) =0$.
This shows that the spectral density $S(\omega)= \lim_{T \to \infty}S_T^{\text{TI}}(\omega)$ is only sensitive to TI part, which is only a function of the squeezing strength
$r(p)$ and not the squeezing angle $\theta(p)$. 
Note that this result does not mean that the angle-sensitive contribution
corresponding to the NTI terms is unphysical. It is only our
chosen observable that turns out to be insensitive to it. An alternative
observable that measures the phase-sensitive part as well is the two-frequency
spectral density $S(\omega, \omega') = \lim_{T \to \infty}\frac{1}{||w_T||^2}
\langle \tilde{x}_T(\omega) \tilde{x}_T^\dagger(\omega')\rangle$.

To summarize this means that the fluctuation scalar which enters the description of the spectral density takes a form analogous to the thermal case
(\ref{eqn:Dhb})
\begin{equation}
	\DTI(p) = \pi \kappa^2 \Sigma^2_n(p)\bigg( \delta_+(p^2) + \delta(p^2)N(p)\bigg) 
\end{equation}
The precise form of the spectrum depends on the momentum-dependence of
the squeezing parameters $r(p)$. Let us consider the simple case
when the squeezing parameter is a function of the energy only $r(p) = r(E_{\vec{p}})$ and so $N(p) =
\sinh^2 r(E_{\vec{p}})$ and we get 
\begin{subequations}
	\begin{align}
		S[\DTI](\omega) 
		&= L \kappa^2  \bigg( \theta(x) F_h(x) + \sinh^2 r(\omega) F_h(|x|) \bigg)
	\end{align}
\end{subequations}

We see that squeezing provides the possibility of exponential enhancement of the
noise amplitude as a function of the squeezing parameter $r(\omega)$.
Generically, squeezing appears in QFT when the background spacetime or the
matter sources are time-dependent \cite{Das:2025kyn}. For instance, squeezing of
states naturally appears in quantum cosmology (inflation)
\cite{grishchuk1990squeezed, albrecht_inflation_1994, Kanno:2025fpz}.  In
quantum gravity we also have squeezing due to the choice of the time evolution
parameter. Such squeezing appears for instance in Unruh states and in the
physics of black holes \cite{hawking1975particle, unruh1976notes}. In these
cases the squeezing is thermal, which means that $\tanh r(\omega) =
e^{-\frac{\pi \omega}{a}}$ with $a$ the observer acceleration in the Unruh case.
More generally, such geometrical squeezing appears when one restricts the study
of QFT to subregions. For instance in a CFT  the restriction of the state to a
spherical region of size $R$ amounts to a thermal state of temperature $ 2\pi/R$
\cite{casini2011towards}.  Note that recent works have already argued that
squeezing could be a window into quantum gravitational effects through its
fluctuation enhancements \cite{Parikh:2020kfh, Guerreiro:2019vbq,
Kanno:2025how,Dorlis:2025zzz, Dorlis:2025amf} similarly
to our calculation, or through the stress tensor renormalization
\cite{Perez:2025tvg}.  The question of whether there exists a realistic
situation where the squeezing parameter becomes large enough for observation is
left to future work.

In this section, we presented detailed derivations of the noise spectral density
for intrinsic fluctuations for the vacuum as well as the thermal and squeezed
states of the canonically quantized linear metric perturbations. We have seen
that the   shape of the homogeneous spectrum is universally determined by
the function $F_h(x)$. Moreover, different
states can introduce suppression or amplification.  For the vacuum state, the
magnitude of the noise spectrum is fully determined by the Planck scale, and is
predicted to be too small for observation in accordance with
\cite{Carney:2024wnp}. For the thermal state, in order for $S(\omega)$ to be of similar magnitude  than the one observed
by interferometers such as LIGO one would need  a
Planckian temperature which is unrealistic.

%% file: sections/sec4_induced.tex
\section{Induced fluctuations}
\label{sec:4}
\newcommand{\DHv}[0]{\Delta^H_{\ket{0}}}
\newcommand{\Hb}[0]{\bar{H}}

In this section, we compute and analyze the noise spectrum arising from
induced fluctuations. Recall from Section \ref{sec:2.QF} that at linear order,
the metric perturbations decompose into intrinsic graviton modes $h^H$, and
induced fluctuations, $H$, sourced by stress-energy backreaction. The
importance of such induced contributions has been emphasized in
\cite{li_interferometer_2023}, where Zurek et al. proposed the pixellon model as
an effective description of non-linear backreaction effects. Here, we use our
framework to access these backreaction effects directly.
We consider a model in which the matter source is a massless scalar quantum
field and compute the noise spectrum for the matter vacuum
state. A detailed analysis of the backreaction from thermal and squeezed states is postponed to future work. We further argue that the present analysis extends to
other stress–energy tensors through the evaluation of two Lorentz-scalars,
including the case of gravitational self-energy. Finally, we compare the intrinsic and induced fluctuation spectra and analyze the UV behavior.

\subsection{Backreaction from massless scalar field}

Let us consider a massless scalar field $\varphi$ that sources the metric
perturbation through the linearized Einstein equation (\ref{eqn:Eequation}). 
\begin{align}
    \Box \varphi = 0, \qquad
		\Box \bar{h}_{ab} = -\frac{\kappa}{2}T_{ab}[\varphi, \eta]
    \label{eqn:HbT}
\end{align}
with the stress tensor\footnote{We use the convention $T_{ab}:=
\frac{-2}{\sqrt{-g}}\frac{\delta S_m}{\delta g^{ab}}\big|_{g=\eta}$ and $S_m =
-\frac{1}{2}\int \rd^4x \, \sqrt{-g} \, (\partial \varphi)^2$}
\begin{equation}
		T_{ab}[\varphi, \eta] = (\partial_a \varphi) (\partial_b \varphi) - \frac{1}{2}\eta_{ab}(\partial \varphi)^2
    \label{eqn:Tphi}
\end{equation}

Using the split $h = h^H + \kappa H$ from (\ref{eqn:homret}), and taking the
trace-reverse, our sourced equation is 
\begin{equation}
    \Box H_{ab} = -\frac{1}{2}\bar{T}_{ab}
\end{equation}
with the trace-reversed stress tensor $\bar{T}_{ab} = (\partial_a \varphi)
(\partial_b \varphi)$, and by definition, $H_{ab}$ is the retarded response
\begin{equation}
    H_{ab}(x) = \frac{1}{2}\int\rd^4y \,G_R(x-y) \bar{T}_{ab}(y)
		\label{eqn:Hprop}
\end{equation}
with the retarded Green's function
\begin{equation}
   G_{R}(x-y) 
    = \frac{1}{2\pi}\delta_{+}[(x-y)^2]
\end{equation}
where $\delta_+(x^2) := \theta(x^0)\delta(x^2)$. In order to calculate the
correlation function $\langle H_{ab} H_{cd} \rangle$,
$H$ needs to be an operator on a
Hilbert space. In this model, (\ref{eqn:Hprop}) means that $H$ lives entirely
in the Hilbert space of the scalar field through the stress tensor, and the
backreaction is mediated via the classical Green's function. This is analogous
to the quantum-controlled field construction from \cite{Telali:2024wcq}.

The scalar field has the mode expansion
\begin{equation}
    \varphi(x) 
    = \int \measure{p} 
    \bigg(
			\ann{p} \, e^{-i p\cdot x} + \text{h.c.}
    \bigg)\bigg|_{p^0 = \Ep{p}}
    \label{eqn:modeexp}
\end{equation}
and is canonically quantized with 
\begin{equation}
    [\ann{p}, a^\dagger_{\vec{p}'}] = (2\pi)^3\, \delta^{(3)}(\vec{p}-\vec{p}')
    \label{eqn:aadagger}
\end{equation}
We denote the momentum space vacuum Wightman function as
\begin{equation}
    \Gv := \int \frac{\rd^4 p'}{(2\pi)^4} 
        \langle \tilde{H}_{ab}(p) \tilde{H}_{cd}(p') \rangle
\end{equation}
which we then write as the convolution with the Green's function
\begin{align}
	\Gv &:= \frac{1}{4}\int \rd^4 \bar{x}\, e^{ip\bar{x}}\int \rd^4 y \int\rd^4 y'
	\\
			& \qquad \times G_R(x-y) G_R(x'-y') \langle \bar{T}_{ab}(y)\bar{T}_{cd}(y')\rangle
			\nonumber
				\label{eqn:Gv}
\end{align}
where $\bar{x} \equiv x-x'$ and the quantum stress tensor is defined via normal
ordering.  Recall that in order to compute the spectral density, we need to
evaluate the contraction $\Sigma^{ab}_n\Sigma^{cd}_n \Gv$ which requires us to
write $\Gv$ in terms of $p_a,\, \eta_{ab}$. In the following, we evaluate the
correlation function for the vacuum state of the scalar field.

\subsubsection{Vacuum state}
The vacuum state of the scalar field $\varphi$ is defined as usual by
$\ann{p}\ket{0} = 0$.  To evaluate the vacuum correlation function $\Gv$ we
first write the correlation function of the trace-reversed stress tensor as
\begin{align}
		&\langle \bar{T}_{ab}(y) \bar{T}_{cd}(y')\rangle_0 \\
		&= \partial_a \partial_c  W^{(0)}(y-y') 
      \partial_b \partial_d  W^{(0)}(y-y') 
    + (c \leftrightarrow d) \nonumber
    \label{eqn:Tbarphi}
\end{align}
which simply follows from the mode expansion (\ref{eqn:modeexp}) and the normal ordering
prescription for the stress tensor (\ref{eqn:Tphi}). Here, $W^{(0)}(y-y')$ is
the vacuum Wightman function of the scalar field
\begin{equation}
    W^{(0)}(y-y') = \int \frac{\rd^4 p }{(2\pi)^3} \delta_+(p^2)e^{-ip(y-y')}
\end{equation}
We can then evaluate the convolution as well as the Fourier transform to get
$\Gv$ by evaluating the Dirac deltas, as detailed
in appendix \ref{app:C.1}. We arrive at the integral 
\begin{equation}
    \Gv =
		\int \frac{\rd \Phi_p(k) }{{4(2\pi)^2(p^2)^2}}
        \bigg(s_a s_c k_b k_d + (c \leftrightarrow d) \bigg)\bigg|_{s = p-k}
    \label{eqn:T0}
\end{equation}
with the measure
\begin{equation}
    \rd\Phi_p(k) := \rd^4 k \, 
        \delta_+(k^2)\, \delta_+((p-k)^2)
\end{equation}
These integrals are the key building block for our analysis. They are commonly
referred to as two-particle Lorentz-invariant phase space integrals
\cite{Schwartz:2014sze} and have already been analyzed in the context of noise
kernels \cite{martin_stochastic_2000}. We provide a comprehensive review of
their structure relevant to our case.

The support of (\ref{eqn:T0}) can be understood by looking at the scalar
building block
\begin{equation}
    I(p) := \int\rd\Phi_p(k) = \frac{\pi}{2} \theta_{+}(p),
    \label{eqn:Ip}
\end{equation}
where $\theta_{+}(p) := \theta(-p\cdot t)\theta(-p^2)$ is the projector onto future-directed timelike vectors. The evaluation of this and all the following
integrals can be found in appendix \ref{app:C}. The main building block needed
for writing the final correlation function is the following integral with a
general Lorentz covariant function $F_{abc...}(k,s)$
\begin{align}
    I^{(F)}{}_{abc...}(p) 
    &:= \int \rd \Phi_p(k)\, F_{abc...}(k,p-k) \nonumber\\
    &= I(p) \, \langle F_{abc...}\rangle(p)
\end{align}
where the spherical average is defined as
\begin{align}
		&\langle F_{abc...}\rangle(p) \\
		&:= \frac{1}{4\pi}\int_{S^2} \rd\Omega(n)\, F(k=\alpha_p(u + n), s =
		\alpha_p(u - n)) \nonumber
\end{align}
with $\alpha_p = \frac{\sqrt{-p^2}}{2},\, u^a = \frac{p^a}{2\alpha_p}, \, n^2 =
1,\, u\cdot n = 0$. This angular average implies that any tensor polynomial of
the vectors $k$ and $s$ will give an expression in terms of the unit momentum
$u^a$ as well as its corresponding transverse projector $P_{ab}:= \eta_{ab} +
u_a u_b$. As detailed in appendix \ref{app:C.2}, our integral takes the form 
\begin{align}
		&\int \rd \Phi_p(k) 
		\bigg(s_a s_c k_b k_d + (c \leftrightarrow d) \bigg) \\
		&= 2\alpha_p^4 I(p) 
				\bigg[
				 u_a u_b u_c u_d -
	\frac{1}{3} 
	\bigg( 
		P_{ab} u_c u_d + P_{cd} u_a u_b
	\bigg) \nonumber \\
	&\quad + \frac{1}{15}
	\bigg(
    P_{ab}P_{cd}+ P_{ac}P_{bd} + P_{ad} P_{bc} 
 \bigg) \nonumber
	\bigg]
\end{align}
where $\alpha_p^4 = (p^2)^2/16$. The final form of the correlation function is
\begin{align}
		&\Gv  \label{eqn:GG} \\
		&=\frac{I(p)}{32(2\pi)^2}
    \bigg[
        u_a u_b u_c u_d - 
        \frac{1}{3}
        \bigg(
            P_{ab}u_c u_d  + P_{cd}u_a u_b 
        \bigg) \nonumber \\
		&+\frac{1}{15}
        \bigg(
            P_{ab}P_{cd}+ P_{ac}P_{bd} + P_{ad} P_{bc} 
        \bigg) \nonumber
    \bigg]
\end{align}
We can check that it satisfies the identity
\begin{equation}
    p^b \Gv 
    = \frac{1}{2}p_a \Gvv{}^b{}_{bcd}(p)
\end{equation}
which is a consequence of energy conservation $\partial^a T_{ab} = 0$.
Equivalently, we can also write the trace-reversed correlation function $\Gbv$,
that is, the correlation function of the trace reversed metric $\bar{H}_{ab}$
using $\langle \bar{H}_{ab} \bar{H}_{cd}\rangle = \langle
(H_{ab}-\frac{1}{2}\eta_{ab}H) (H_{cd}-\frac{1}{2}\eta_{cd}H)\rangle$,
which takes a simpler form
\begin{equation}
    \Gbv
		= \frac{I(p)}{80(2\pi)^2} 
    \bigg[
        P_{ab}P_{cd} + \frac{1}{6}
        \bigg(
            P_{ad}P_{bc} + P_{ac}P_{bd}
        \bigg) 
    \bigg] 
		\label{eqn:GbGb}
\end{equation} 
Energy conservation is manifest in this form 
\begin{equation}
    p^a \Gbv = 0
		\label{eqn:cons}
\end{equation}
since it is determined from $\langle T_{ab} T_{cd} \rangle$ and thus only
composed of the transvere projectors and then is trivially satisfied. 

\subsubsection{General correlation functions}
\newcommand{\Gbab}[0]{\bar{\mathcal{G}}^{(A,B)}_{abcd}(p)}
\newcommand{\Gab}[0]{\mathcal{G}^{(A,B)}_{abcd}(p)}
\newcommand{\Gabi}[0]{\mathcal{G}^{(A,B)}}
\newcommand{\Gbabi}[0]{\bar{\mathcal{G}}^{(A,B)}}
We note that it is possible to write the general solution of the conservation
equation (\ref{eqn:cons}) as\footnote{The convention with the factor of
$\frac{1}{4}$ is used to make the relation to the stress-energy correlator
simpler.}
\begin{equation}
        \Gbab:= \frac{1}{4}\bigg( A(p) P_{ab}P_{cd} + B(p) P_{abcd} \bigg)
        \label{eqn:ab}
\end{equation}
where $A,B$ are Lorentz scalars of the momentum $p$ and 
\be
P_{abcd} = \frac{1}{2}P_{ac}P_{bd} + \frac{1}{2}P_{ad}P_{bc} - \frac{1}{3}P_{ab}P_{cd}
\ee
 is the symmetric transverse and  traceless projector, which satisfies
 $\eta^{ab}P_{abcd} = p^a P_{abcd}= 0$.  $\Gbab$ is the most general tensor that
 satisfies transversality, Lorentz covariance, and the symmetry properties
\begin{equation}
    \Gbabi_{abcd} = \Gbabi_{cdab} = \Gbabi_{bacd}
\end{equation}
To calculate the trace reversed form $\Gab$, we need
\begin{align}
	&\eta^{cd}\, \Gbab = \frac{3A}{4} P_{ab}\\
	&\eta^{ab}\eta^{cd}\, \Gbab = \frac{9A}{4} \nonumber
\end{align}
and then we have
\begin{align}
	&\Gabi_{abcd} \\
	&= \frac{1}{16}\bigg( A(p) P_{ab}P_{cd}+ 4 \,B(p) P_{abcd} + 9A(p) u_a u_b u_c
	u_d \nonumber \\
	&\quad - 3A(p)\big( P_{ab}u_c u_d + P_{cd}u_au_b\big)\bigg) \nonumber
    \label{eqn:Wcoorab}
\end{align}
Our results for the scalar field (\ref{eqn:GbGb}) and (\ref{eqn:GG}) are
retrieved by setting
\begin{equation}
    A(p) = \frac{I(p)}{18(2\pi)^2}, \qquad B(p) = \frac{3A(p)}{10}
\end{equation}
In general, $A$ and $B$ will be proportional to $I(p)$ with coefficients that depend on the mass and spin of the field.

\subsection{Vacuum spectral density}
We now proceed to calculate the spectral
density. We need to evaluate the fluctuation scalar (\ref{eqn:Deltapp}) which
for the sourced metric $H$ is
\begin{equation}
	\DHv(p) := \kappa^4 \Sigma^{ab}_n \Sigma^{cd}_n \Gv
\end{equation}

where $\Gv$ is given by (\ref{eqn:GG}). Similarly to the
homogeneous case, gauge invariance $\Sigma^{ab}_n p_b = 0$ implies that it is
only the background metric $\eta_{ab}$ in the transverse projector $P_{ab}$
that contributes. This means that apart from numerical factors, the main difference between the fluctuation
scalar of the homogeneous perturbation from (\ref{eqn:Dhv}) and the sourced
model is the scalar function $I(p)$. Recall that $I(p)$
has support on the inside of the future light cone as
opposed to $\delta_+(p^2)$ for the homogeneous perturbation, which has support
on the future light cone. The fluctuations scalar takes the form  
\begin{align}
     \DHv(p) &=  \frac{\kappa^4\theta_{+}(p)}{5120 \pi}\bigg( 1- \frac{(p\cdot n)^2}{(p\cdot t)^2}\bigg)^2
		\label{eqn:DHv}
\end{align}
Plugging this propagator into the generic formula (\ref{eqn:S}) for the spectral
density, we get the following spectrum
\begin{equation}
    S[\DHv](\omega) = \frac{\kappa^4}{L}\theta(x) F_H(x)
		\label{eqn:SDH}
\end{equation}
where we again introduced the dimensionless quantity $x:= \omega L$ and
the dimensionless function
\begin{align}
	F_{H}(x) &:=\frac{x}{160(2\pi)^3}\bigg[ \frac{2}{15} +\frac{1}{x^2} -
		\frac{15}{2x^4} - \cos(2x)\bigg(\frac{1}{15} + \frac{9}{2x^4}\bigg)
		\nonumber \\
	&\qquad + \sin(2x)\bigg( \frac{6}{x^5}- \frac{1}{x^3}\bigg)  \bigg] 
		\label{eqn:FH}
\end{align}
which we plot in figure \ref{fig:SH} together with its enveloping curve.  The
evaluation is detailed in appendix \ref{app:B.3}. 

\begin{figure}[h!]
    \centering
    \includegraphics[width=1.0\linewidth]{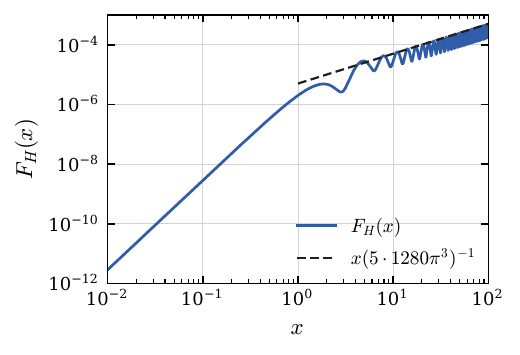}
			\caption{Dimensionless function $F_H(x)$ determining the induced vacuum
			spectrum $S[\DHv](\omega)=\kappa^4 L^{-1}\theta(x)F_H(x)$, with $x=\omega
			L$. The dashed curve shows the large-$x$ upper envelope
			$x/(5\cdot1280\pi^3)$.}
    \label{fig:SH}
\end{figure}

\paragraph*{Behavior in the limits.}
In the expansion at $x=0$ all the singular terms cancel, the first non-zero
term is of order $x^3$ and given by 
\begin{equation}
	F_H(x)= \frac{x^3}{11200 \pi^3} + O(x^5) .
\end{equation}
At large $x$, one finds that
\begin{align}
	F_H(x)
	  &=\frac{x}{1280\pi^3}\bigg[\frac1{15} \big(1+ 2 \sin^2x\big)
		+\frac{1}{x} \nonumber \\
		&\quad -\frac{\sin 2x}{x^2}
		+\cO\left(\frac{1}{x^3}\right) \bigg].
\end{align}
The dominant oscillation comes from the $\cos(2x)$. The leading upper and lower
envelopes are therefore
\begin{subequations}
\begin{align}
	F_H^{\max}(x) &\sim \frac{x}{6400\pi^3}, \\
	F_H^{\min}(x) &\sim \frac{x}{19200\pi^3}.
\end{align}
\end{subequations}
 with subleading
$1/x$-suppressed oscillatory corrections.
\paragraph*{Total power.}
The total power in the spectrum has a quadratic UV-divergence
\begin{equation}
    \int_0^\Lambda \rd \omega\,S[\DHv](\omega) \sim \kappa^4 \Lambda^2
\end{equation}
Recall from \eqref{eqn:powerSh} that the intrinsic spectrum has logarithmically
UV-divergent total power.
\paragraph*{Magnitude of the noise.}
In natural units, the spectrum has magnitude
\begin{align}
    S[\DHv](\omega)  
    \sim \bigg( 
    10^{-72}\sqrt{F_{H}\big(2\pi f L\big)}
    \frac{\text{m}}{\sqrt{\text{Hz}}}\bigg)^2
    \bigg(\frac{1 \,\text{m}}{L}\bigg) 
\end{align}
Since the envelope of $F_{H}$ is monotonically increasing, the signal gets
stronger with the frequency. For illustrative purposes, we can determine that we
need a frequency $f \sim 10^{120}\, \text{Hz}$ in order to make the signal as
strong as $10^{-19} \frac{\text{m}}{\sqrt{\text{Hz}}}$, which is the amplitude of
the LIGO noise \cite{buikema_sensitivity_2020}.

\subsection{Comparison of intrinsic and induced spectra}
Let us now compare the spectra of the intrinsic \eqref{eqn:SDh2} and induced fluctuations \eqref{eqn:SDH}.
One has that 
\be 
\frac{S[\DHv](\omega)}{S[\Dhv](\omega)}
= (\kappa \omega)^2
\frac{ F_H(x)}{x^2  F_h(x)}
\ee
At low and large frequency, this ratio becomes 
\begin{subequations}
    \begin{align}
    \frac{S[\DHv](\omega)}{S[\Dhv](\omega)}
    & = 
    \frac{3}{280} \left(\frac{\kappa \omega}{4\pi}\right)^2 + O(x),\\
     & = \frac{ 1 }{100} \left(\frac{\kappa \omega}{4\pi}\right)^2
		 \frac{\big(1+  2\sin^2x\big)}{\left(1 + \sin^2 x\right)} + O(x^{-1})
		 \nonumber\\
		 &\lesssim \frac{ 3 }{200} \left(\frac{\kappa \omega}{4\pi}\right)^2
		 \nonumber
     \end{align}
\end{subequations}
 Since $\kappa=\sqrt{32\pi \ell_P^2}$, we see that the induced fluctuation
 spectrum is suppressed compared to the intrinsic spectrum. The large $\omega$
 behavior here means large with respect to $L^{-1}$ but small compared to the
 Planck scale  $L^{-1}\ll\omega \ll\ell_P^{-1}$. This is in agreement with the
 expected EFT scaling where higher order terms in the perturbation series are of
 order $\hbar G_N \omega^2 =E^2/M_{\text{Pl}}^2 $ where $\omega=E/\hbar$ is the
 wave number and $E$ is the process energy \cite{Donoghue:2012zc}. This means
 that the increase of the induced fluctuation relative to the intrinsic one is
 connected to the non-renormalizability of gravity. 

\paragraph*{Discussion of short-distance singularities}

We now clarify how the power spectral density encodes the short-distance singularities of the underlying quantum field theory. As we have
established in \eqref{eqn:timedelayFkernel}, the proper time fluctuation $\delta
\tilde{\tau}(\omega)$ is related to the metric via the response function $R(\omega,
\vec{p})$, which provides the smearing of the Wightman function to give the
spectral density as shown in \eqref{eqn:SG}. To isolate the essential
mechanism, it is instructive to consider simplified position-space examples in
which the role of the Wightman function can be made explicit. 

Consider a schematic 'bare' spectrum given by the Fourier transform of the unsmeared vacuum Wightman function of a free scalar field $\varphi$
\begin{equation}
    S(\omega) = \int_{-\infty}^{\infty} \rd t \, e^{-i\omega t}\, W(t),
    \label{eqn:nakedS}
\end{equation}
where $W(t) := \bra{0}\varphi(t, \vec{x} = 0)\varphi(0)\ket{0}$. The position-space form of $W$ may be written as
\begin{align}
	W(t) &= \int \frac{\rd^4 p}{(2\pi)^3}\,\delta_+(p^2)\,e^{i p^0 t} \\
			 &= \frac{1}{(2\pi)^2}\,\frac{-1}{(t+i\epsilon)^2} \nonumber,
    \label{eqn:Wt}
\end{align}
where the $i\epsilon$ prescription defines it as a distribution.
The Fourier transform in \eqref{eqn:nakedS} is therefore determined by the
residue at the pole $t_*=-i\epsilon$
\begin{align}
    S(\omega)
		&= -2\pi i\,\Res\!\left(e^{-i\omega t}W(t),\,t_*\right) \\
		&= \lim_{\epsilon\to 0} e^{-\epsilon\omega}\,\frac{\omega}{2\pi} \nonumber
    = \frac{\omega}{2\pi},
\end{align}
where the minus sign comes from closing the contour in the lower half-plane.
This spectrum exhibits a linear growth in the frequency, in direct analogy with
the induced fluctuation spectrum \eqref{eqn:SDH}. From this simplified example,
it is clear that this behavior reflects the Hadamard short-distance singularity
$W(t)\sim t^{-2}$ of the free vacuum.

Next, consider a sourced field whose Wightman function is supported
inside the future light cone in momentum space. Schematically, we take
\begin{equation}
    W_J(t)
    := \int \rd^4p \, e^{i p^0 t}\, I(p),
\end{equation}
where the function $I(p) = \frac{\pi}{2}\theta_+(p)$ was defined in \eqref{eqn:Ip}. This models the Wightman function of induced fluctuations from \eqref{eqn:GG}.
Carrying out the momentum integral explicitly gives
\begin{align}
    W_J(t)
		&= 2\pi^2\int_0^{\infty} \rd p^0\, e^{i p^0 t}
      \int_0^{p^0} \rd |\vec p|\, |\vec p|^2 \\
		&= \frac{2\pi^2}{3}\int_0^{\infty} \rd p^0\, e^{i p^0 t}\,(p^0)^3
		= \frac{(2\pi)^2}{(t+i\epsilon)^4} \nonumber
\end{align}
This function exhibits a fourth-order short-distance
singularity, stronger than the Hadamard singularity of the homogeneous free field \eqref{eqn:Wt}.
The corresponding spectrum is again obtained from the residue at
$t_*=-i\epsilon$,
\begin{align}
    S(\omega)
		&= -2\pi i\,\Res\!\left(e^{-i\omega t}W_J(t),\,t_*\right) \\
		&= \lim_{\epsilon\to 0} e^{-\epsilon\omega}\,\frac{4\pi^3}{3}\,\omega^3
		\nonumber
    = \frac{4\pi^3}{3}\,\omega^3,
\end{align}
which grows cubically with the frequency, reflecting the higher-order short-distance divergence.

These examples illustrate a general pattern. If the short-distance
limit of a Wightman function has the scaling
\begin{equation}
    W(t) \sim \frac{1}{(t+i\epsilon)^{2n}},
\end{equation}
then its unsmeared Fourier transform scales as $S(\omega)\sim \omega^{2n-1}$.
If we now take the smearing with the response function $R$ into account, we see from \eqref{eqn:SG}, that $|R|^2$ contributes with a
suppression factor of $\sim \omega^{-2}$. Overall, the net scaling of the measured spectrum of a Wightman function with order $2n$ short-distance singularity is 
\begin{equation}
    S(\omega)\sim \omega^{2n-3}.
\end{equation}
Accordingly, intrinsic vacuum fluctuations with $n=1$ lead to a $1/\omega$
falloff as observed in \eqref{eqn:SDh2}, while induced fluctuations with $n=2$
exhibit a linear rise \eqref{eqn:SDH}. The scaling behavior of the spectrum
therefore directly encodes the order of the short-distance singularity of the
underlying Wightman function once the measurement kernel is taken into account.

%% file: sections/conclusion.tex
\section{Conclusion}

\label{sec:conclusion}

In this work, we studied the power spectral density (PSD) of interferometric time delay fluctuations caused by space-time fluctuations as part of a growing effort to identify an observable signature of quantum gravity in low-energy experiments. We derived a general gauge-invariant formula for the PSD for any
translation-invariant state of the linear metric perturbation around Minkowski spacetime. It takes the scalar
contraction of the metric Wightman function as its input, and can be used to evaluate many different cases within one framework. 

We first applied this formalism to evaluate the effect of homogeneous
fluctuations in the vacuum, thermal, and squeezed states of the graviton field.
For the vacuum state we reproduced the result of Carney et al.
\cite{Carney:2024wnp}, namely that the noise is fully determined by the Planck
scale and it is too small for observation. The thermal and squeezed states
provide mechanisms for enhancing the fluctuations compared to the vacuum
spectrum.  In the thermal case the graviton enhancement does not appear
realistic as an observable magnitude requires Planckian temperature. For the
squeezed state, whether an observable enhancement is realistic is still open.
All these homogeneous spectra oscillate and decay inversely with the frequency.

We next presented the evaluation of the effects of fluctuations induced by the stress-energy of a massless scalar field, which constitutes one of the main results of the paper.
We find a spectrum that oscillates, has a linearly growing envelope at high frequencies, and it is suppressed relative to the homogeneous
contribution by an overall factor of $(\kappa \omega)^2$.  
Crucially, we find no problematic UV divergences in the two-point function of the stress
tensor, as the convolution with the retarded propagators smoothens the singular
behavior just enough to render the result finite.
Therefore, our analysis does not indicate any breakdown to the EFT paradigm within the
scope of the observable considered here. This breakdown is necessary to allow
for an observable signature of the Verlinde-Zurek effect \cite{Verlinde:2019xfb}.

However, our analysis is not complete, and several theoretical possibilities
remain unexplored, and so we cannot yet provide a definitive conclusion. First,
for a complete treatment of backreaction, one must include the stress-energy
tensor of the first-order metric fluctuations as a source for the induced
metric. At this order, the consistency of the metric backreaction requires the
inclusion of the second-order geodesic deviation, which was neglected in this
work.  Second, the backreaction of massive fields may exhibit a richer structure
arising from the analytic continuation of the spectral functions into the
complex frequency plane, as suggested by \cite{Afshordi:2015iza,
Afshordi:2017scc, Afshordi:2019xbz}, making this another interesting direction
to pursue. Furthermore, while we focused on a single interferometer arm, a key
observable proposed by Zurek and Verlinde and developed in the context of the
pixellon model \cite{li_interferometer_2023,
Verlinde:2022hhs, Verlinde:2019xfb} involves angular
correlations between two different arms, and thereby probing the quantum
geometry of the causal diamond. In this setting, we expect a universal response
function controlling fluctuations.

In addition, our analysis does not yet account for the IR cutoff that enters
through the finite time of observation. Indeed, the conjectured observability of
fluctuations is related to UV/IR mixing, and it is therefore necessary to assess
whether this mixing is related to the IR cutoff entering the design of the
observable. Finally, in quantum field theory, the notion of the Fock vacuum
requires a choice of reference frame, breaking covariance. In quantum gravity,
however, one needs to include a choice of quantum reference frame (QRF), which
is itself subject to quantum fluctuations. This QRF enters the construction of a
covariant Fock vacuum and the normal ordering prescription which defines the
stress-tensor \cite{Ciambelli:2023mir, Ciambelli:2024swv,
Freidel:2025ous, Freidel:2026stu}. One then needs to assess the effect of the QRF
fluctuations on the final observable, which is one of the most promising avenues
in our view. 

A related effect arises from the renormalization of the stress tensor two-point
function. The renormalization, needed to preserve covariance, can affect the
consistency condition, which requires that stress tensor fluctuations be smaller
than their expectation values. If this is violated, the induced fluctuation
dominates the noise spectrum opening up the possibility of observable
fluctuations. It has recently been argued in \cite{Perez:2025tvg,
Freidel:2026stu} that such a violation occurs for squeezed states due to
renormalization effects and to fluctuations of the quantum reference frames.
One needs to assess whether this can affect the time delay PSD.

%% file: appendices/appA_time_delay.tex
\section{Time delay calculation}
\label{app:A}
In this appendix, we detail the derivation of the time delay formula (\ref{eqn:timedelay1}). We start with the geodesic equation for a curve $x^a(\lambda)$ 
\begin{equation}
    \frac{\rd^2 x^a(\lambda)}{\rd \lambda^2} + \Gamma^a{}_{bc}(x(\lambda)) \frac{\rd x^b(\lambda)}{\rd \lambda}\frac{\rd x^c(\lambda)}{\rd \lambda} = 0
\end{equation}
with affine parameter $\lambda$. Let us define the time parameter $\tau$ as 
\begin{equation}
    \tau := -x^a(\lambda)t_a
\end{equation}
and denote the derivative with respect to $\tau$ as $\dot{x}^a \equiv \frac{\rd x^a}{\rd\tau}$. This parametrization implies
\begin{equation}
    \dot{x}^a t_a = -1.
\end{equation}
 Using the chain rule
\begin{subequations}
\begin{align}
    \frac{\rd x^a(\lambda)}{\rd \lambda} &= \dot{x}^a\frac{\rd \tau}{\rd \lambda},\qquad 
    \frac{\rd^2 x^a(\lambda)}{\rd \lambda^2} = \ddot{x}^a\bigg(\frac{\rd \tau}{\rd \lambda} \bigg)^2 + \dot{x}^a \frac{\rd^2 \tau}{\rd \lambda^2}
\end{align}
\end{subequations}
the geodesic equation becomes
\begin{equation}
\label{geodeq}
   \ddot{x}^a(\tau) +  \Gamma^a(\tau) =0
    \qquad 
   \Gamma^a(\tau) := 
   \dot{x}^b \dot{x}^c \bigg( \Gamma^a{}_{bc} + \dot{x}^a\Gamma^t{}_{bc}\bigg)(x(\tau)).
\end{equation}
where we used that the contraction of the equation with $t_a$, and $\Gamma^a$ is such that  $t_a \Gamma^a =0$. In the temporal gauge  defined by
\begin{equation}
    t^a h^{(t)}_{ab} = 0
\end{equation}
we have that $\Gamma^a{}_{tt} = 0$. 
This implies
that a geodesic that starts with velocity $\dot{x}^a(0)=t^a$ stays parallel to $t^a$. For such geodesics, 
$\tau$ is the proper time and 
 their geodesic motion is simply given,
at all order in $\kappa_c$, by
\begin{equation}
   x^a(\tau)  = (\tau - \tau_0) \,t^a + x_0^a.
\end{equation}
These represent the geodesic motion of the mirror and beamsplitter in temporal gauge.

For a null geodesic, we obtain corrections from the metric perturbation. Let us look for generic solutions in the form
\begin{equation}
    x^a(\tau) = \ell^a(\tau-\tau_0) + x_0^a + \delta x^a(\tau)
\end{equation}
where $\ell$ is a null vector $\ell^2=0$ such that $\ell \cdot t = -1$, and $\delta x^a(\tau)$ is an order $\kappa_c$ correction. Note that we parametrize the null geodesics with $\tau$ which is not an affine parameter for the perturbed geodesic but it is convenient since it agrees with the proper time of the timelike curves at the intersection points. The tangent vector is simply
\begin{equation}
    \dot{x}^a = \ell^a + \delta \dot{x}^a
\end{equation}
The condition that $\tau$ is the timelike geodesic proper time implies that 
\be 
\delta \dot{x}^a t_a =0.
\ee
On the other hand 
the condition that the tangent vector is null becomes
\begin{equation}
    (\eta + \kappa_c h_{ab})(\ell^a + \delta \dot{x}^a)(\ell^b + \delta \dot{x}^b) = \kappa_c h_{\ell \ell} + 2\ell \cdot \delta \dot{x} = 0
\end{equation}
that is
\begin{equation}
    \ell \cdot \delta\dot{x} = -\frac{\kappa_c}{2}h_{\ell\ell} 
\end{equation}
This means that the perturbation can be decomposed as 
\begin{equation}
    \delta \dot{x}(\tau) = -\alpha(\tau) n^a + \delta \dot{x}^a_\perp \qquad 
    \alpha(\tau) : =  \frac{\kappa_c}{2}h_{\ell\ell}(x(\tau))
\end{equation}
where $n$ is the spacelike direction of the null geodesic, i.e $\ell= t +n$ and  
where $\delta x^a_\perp$ is the geodesic fluctuation perpendicular to $t$ and $n$.
The acceleration coefficient entering the geodesic evolution \eqref{geodeq} can be decomposed in the temporal gauge as 
\be 
\Gamma^a(\tau)=  \pa_\tau \alpha(\tau) n^a  + \Gamma_{\perp}^a(\tau) 
\ee
where $\Gamma_{\perp}^a(\tau)$ is the projection of $\Gamma^a(\tau)$ in the plane orthogonal to $(t,n)$.
This means that the solution of the geodesic equation is, to first order, given by
\be 
\delta x^a(\tau) =  v_\perp^a (\tau-\tau_0) - n^a \int_{\tau_0}^\tau  \alpha(\tau') \rd \tau' 
+  \int_{\tau_0}^\tau (\tau-\tau') \Gamma^a_\perp(\tau') \rd \tau'.
\ee 
where we assume that $\delta x(\tau_0)=0$.
The orthogonal  velocity $v_\perp$ is 
determined by the condition that the geodesic comes back to the $(t,n)$ plane at the arrival time $\tau_1$, so that $\delta x_\perp(\tau_1)=0$. This means that 
\be 
v_\perp^a = -\frac1{\tau_1-\tau_0}   \int_{\tau_0}^{\tau_1} (\tau_1-\tau) \Gamma^a_\perp(\tau) \rd \tau.
\ee 
which fully determines 
$\delta x^a$ in terms of the metric perturbation.
Note that we have that 
\be 
\Gamma_{\perp}^a(\tau) =\kappa_c
 P_\perp^{ab} \left(\pa_\tau[ h_{b \ell}(x(\tau))] -  \frac12 \pa_b h_{\ell\ell} (x(\tau))
 \right)
\ee 
where $P_\perp$ is the projector onto the plane orthogonal to $(t,n)$.

We can now use these generic geodesic curves to write down our solutions for the mirror, beamsplitter and the light rays.
The non-perturbative geodesics need to  satisfy the matching conditions
\begin{subequations}
    \begin{align}
        {x}_B(\tau_E ) &= {x}_+(\tau_E) \\
       {x}_M(\tau_E+L +\delta\tau_+) &= {x}_+(\tau_E+L +\delta\tau_+)= {x}_-(\tau_E+L + \delta\tau_+) \\
        {x}_B(\tau_E + 2L + \delta\tau_++\delta\tau_- ) &= {x}_-(\tau_E+2L +\delta\tau_++\delta\tau_-)
    \end{align}
\end{subequations}
where $\delta \tau_\pm$ are the proper time shift of arrival of the outgoing (resp. incoming ) geodesics and the mirror (M) (resp. the beamsplitter (B)). The zeroth order background geodesics that satisfy these are
\begin{subequations}
    \begin{align}
        \bar{x}_{B,M}(\tau) &= \tau\, t +  x_{B,M} \\
        \bar{x}_{+}(\tau) &= \tau\, t +  (\tau-\tau_E) n + {x}_{B} \\
        \bar{x}_{-}(\tau) &= \tau\, t -   (\tau-\tau_E - 2L) n + x_B
       \label{eqn:geodesics2} 
    \end{align}
\end{subequations}
To linear order and in temporal gauge, where $\delta x_{B,M}=0$, the matching conditions imply that
\begin{subequations}
    \begin{align}
     \delta \tau_+ n^a &=   -(\delta x_+(\tau_E+L)-\delta x_+ (\tau_E) ) = n^a 
     \int_{\tau_E}^{\tau_E + L} \alpha_+(\tau) \rd \tau,\\
     \delta \tau_- n^a &=   \delta x_-(\tau_E+2L)-\delta x_+ (\tau_E+L) = -n^a 
     \int_{\tau_E +L}^{\tau_E + 2 L} \alpha_-(\tau) \rd \tau
    \end{align}
    \label{eqn:dtauna}
\end{subequations}
where $\alpha_\pm(\tau)$ is the contraction of the metric in the temporal gauge along the outgoing and incoming unperturbed tangent null vectors $\ell_\pm = t \pm n$, and is evaluated on the background geodesic
\begin{equation}
    \alpha_\pm(\tau) := \pm \frac{\kappa_c}{2}h^{(t)}_{\ell_\pm \ell_\pm}(\bar{x}_\pm(\tau))
\end{equation}
Contracting (\ref{eqn:dtauna}) with $n^a$ and substituting $\alpha_\pm$, we get the final time delay formulas in the temporal gauge
\begin{subequations}
    \begin{align}
        \delta\tau_+ &= \frac{\kappa_c}{2}\int_{\tau_E}^{\tau_E+L}\rd \tau \, h^{(t)}_{\ell_+ \ell_+}(\bar{x}_+(\tau)), \\
        \delta\tau_- &= \frac{\kappa_c}{2}\int_{\tau_E+L}^{\tau_E+2L}\rd \tau \,h^{(t)}_{\ell_-\ell_-}(\bar{x}_-(\tau)).
    \end{align}
\end{subequations}

%% file: appendices/appB_psd.tex
\section{Power spectral density calculations}
\label{app:B}

In this appendix, we first detail the derivation of the general form
(\ref{eqn:S}) of the power spectral density of a translation-invariant state as
a function of the fluctuation scalar, and then derive the final forms for the
vacuum state in the case of intrinsic and induced fluctuations from equations
(\ref{eqn:SDh}) and (\ref{eqn:SDH}) respectively. Finally, we prove how the infinite-time limit of finite measurement of the spectral density leads to the form \eqref{eqn:Stau} for a translation-invariant state, and discuss what happens to the non-translation-invariant contribution that appears in the squeezed state in Section \ref{sec:3.squeezed}.

\subsection{General form of the power spectrum}
\label{app:B.1}

We evaluate the response function $R = R_+ + R_-$ starting from its definition in
(\ref{eqn:Fkernel}), which we restate here for convenience
\begin{equation}
    2\pi\delta(p^0-\omega) R(\omega, \vec{p}) :=\int_{-\infty}^\infty \rd \tau_E \, e^{-i\omega \tau_E}\bigg(\int_{\tau_E}^{\tau_E + L}\rd \tau \, e^{-i \bar{x}_+(\tau)\cdot p} +  \int_{\tau_E+L}^{\tau_E + 2L}\rd \tau \, e^{-i \bar{x}_-(\tau)\cdot p} \bigg)
\end{equation}
Using the explicit form of the background geodesics $\bar{x}_\pm(\tau)$ from
(\ref{eqn:geodesics}), we can regroup the exponents for the $+$ and $-$ branches as
\begin{equation}
	p\cdot \bar{x}_+(\tau) + \omega \, \tau_E = -(p^0 - \pn)\tau + (\omega- \pn)\tau_E + \vec{p}\cdot \vec{x}_B.
\end{equation}
and
\begin{equation}
	p\cdot \bar{x}_-(\tau) + \omega \,\tau_E = -(p^0 + \pn)\tau + (\omega+ \pn)\tau_E +  2L \vec{p}\cdot \vec{n} + \vec{p}\cdot \vec{x}_B.
\end{equation}
The integrals over the proper time $\tau$ respectively give
\begin{equation}
    \int_{\tau_E}^{\tau_E + L} \rd \tau \, e^{i(p^0 - \pn)\tau} = \frac{e^{i(p^0
		- \pn)L}-1}{i(p^0 - \pn)} e^{i(p^0-\vec{p} \cdot \vec{n})\tau_E}
\end{equation}
and
\begin{equation}
    \int_{\tau_E+L}^{\tau_E + 2L} \rd \tau \, e^{i(p^0 + \pn)\tau} =
		\frac{e^{i(p^0 + \pn)\tau_E}}{i(p^0 + \pn)}\bigg(  e^{i(p^0+\vec{p} \cdot \vec{n})2L} -  e^{i(p^0+\vec{p} \cdot \vec{n})L}\bigg)
\end{equation}
In both cases, the factor $(p\cdot \vec{n}) \tau_E$ cancels in the exponent and the integral over $\tau_E$ gives $2\pi\,\delta (p^0 - \omega)$
which is factored out of the definition of $R$, and hence we get the form
(\ref{eqn:Fkernelfinal}) for the response functions
\begin{subequations}
    \begin{align}
        R_+(\omega, \vec{p}) &= \left(
     \frac{e^{i(\omega  - \pn)L}-1}{i(\omega  - \pn)}
     \right)e^{-i \vec{p}\cdot \vec{x}_B} \\
        R_-(\omega, \vec{p}) &= \bigg(\frac{e^{i(\omega + \pn)2L} -e^{i(\omega +
				\pn)L}}{i(\omega + \pn)}\bigg) e^{-i(\vec{p}\cdot \vec{x}_B + 2L\,\pn)}
    \end{align}
\end{subequations}

We now calculate the general form of the spectral density $S[\Delta]$ from
(\ref{eqn:S}), which, based on (\ref{eqn:SG}), only requires the evaluation of
the modulus of the response function $|R|^2 = |R_+|^2 + |R_-|^2 + R_+ R_-^* + R_- R_+^*$.
For all contributions, the factors containing the position of the beamsplitter
$\vec{x}_B$ cancel. For the $++$ branch, we get
\begin{align}
	|R_+|^2 &= \frac{1}{(\omega - \pn)^2} 
	\bigg( e^{i(\omega - \pn)L} -1 \bigg) 
	\bigg( e^{-i(\omega - \pn)L} -1 \bigg) \nonumber\\
					&= \frac{1}{(\omega - \pn)^2} 
	\bigg(2-2\cos((\omega - \pn)L)\bigg)  \nonumber\\
					&= 4 \bigg( \frac{\sin((\omega-\pn)L/2)}{(\omega - \pn)} \bigg)^2
\end{align}
For the $--$ branch, it is the same calculation except that the end result is
obtained by the replacement $\vec{p} \to -\vec{p}$
\begin{align}
	|R_-|^2 &= \frac{1}{(\omega + \pn)^2} 
	\bigg( e^{i(\omega + \pn)2L} -e^{i(\omega + \pn)L} \bigg) 
	\bigg( e^{-i(\omega + \pn)2L} -e^{-i(\omega + \pn)L} \bigg) \nonumber\\
					&= \frac{1}{(\omega + \pn)^2} 
	\bigg(2-2\cos((\omega + \pn)L)\bigg)  \nonumber\\
					&= 4 \bigg( \frac{\sin((\omega+\pn)L/2)}{(\omega + \pn)} \bigg)^2
\end{align}
For the $+-$ branch
\begin{align}
	R_+ R_-^* &= \frac{1}{\omega^2 - (\pn)^2} 
	\bigg( e^{i(\omega - \pn)L} - 1\bigg)
	\bigg( e^{-i(\omega + \pn)2L} - e^{-i(\omega + \pn)L} \bigg)e^{2iL \pn}
	\nonumber \\
						&= \frac{1}{\omega^2 - (\pn)^2} 
						\bigg( e^{-i(\omega + \pn)L} - e^{-i2\omega L} - 1 + e^{-i(\omega -
						\pn)L}\bigg) \nonumber \\
						&= \frac{2e^{-i\omega L}}{\omega^2 - (\pn)^2} 
						\bigg( \cos(\pn L) - \cos \omega L\bigg) 
\end{align}
Finally, since $R_+ R_-^* = (R_- R_+^*)^*$, the $-+$ branch is simply the
complex conjugate
\begin{equation}
	R_- R_+^* = \frac{2e^{i\omega L}}{\omega^2 - (\pn)^2} 
						\bigg( \cos(\pn L) - \cos \omega L\bigg) 
\end{equation}
Plugging these four contributions into (\ref{eqn:SG}) yields the general
expression for the spectrum $S[\Delta]$ from (\ref{eqn:S}).

\subsection{Intrinsic vacuum spectrum}
\label{app:B.2}
We now compute the final form of the power spectral density (\ref{eqn:SDh}) for the
intrinsic fluctuations in the vacuum state. We start from the general formula
for any translation-invariant state (\ref{eqn:S}) and use the fluctuation scalar
for the vacuum state (\ref{eqn:Dhv}) which we restate with the substitution
$p^a=(\omega, \vec{p})$
\begin{equation}
	\Dhv(\omega, \vec{p}) = \pi \kappa^2 \Sigma_n^2(\omega,
	\vec{p})\theta(\omega)\delta(-\omega^2 + |\vec{p}|^2), \quad 
	\Sigma_n^2(\omega, \vec{p}) = \frac{1}{4} 
	\bigg( 1 - \frac{(\pn)^2}{\omega^2}\bigg)^2
\end{equation}
We compute explicitly the $++$ and $+-$ branches. The $--$ branch is identical
to $++$ under $\vec{p}\to-\vec{p}$ due to spatial translation-invariance, while
the $-+$ branch is the complex conjugate of $+-$.  For the $++$ branch, the full
integral we need to evaluate is
\begin{equation}
	S_{++}[\Dhv](\omega) = \pi \kappa^2 \theta(\omega) \int \frac{\rd^3 \vec{p}}{(2\pi)^3} 
	\left(
		\frac{\sin\left((\omega  - \vec{p}\cdot \vec{n}\right)L/2)}{(\omega  - \vec{p}\cdot \vec{n}) }
  \right)^2 
	\bigg( 1 - \frac{(\pn)^2}{\omega^2}\bigg)^2
	\delta(-\omega ^2 + |\vec{p}|^2)
\end{equation}
Since the integrand only depends on $|\vec{p}|$ and $\pn$, we use polar
coordinates with polar axis along $\vec{n}$ so the measure becomes $\rd^3
\vec{p} = |\vec{p}|^2 \rd |\vec{p}|\, \rd \phi \, \rd u$, where $\phi \in [0,
2\pi]$ is the azimuthal angle and $u := \frac{\pn}{|\vec{p}|} = \cos \theta \in
[-1,1]$ is the cosine of the polar angle $\theta$ between $\vec{p}$ and
$\vec{n}$. The azimuthal integral gives a factor of $2\pi$ as the integrand is
independent of $\phi$, and the integral with respect to $|\vec{p}|$ evaluates
the $\omega \geq 0$ branch\footnote{The $\omega <0$ branch evaluates to the same integral, which is needed for the thermal and squeezed states.} of the Dirac delta which we write as $\delta(-\omega
^2 + |\vec{p}|^2) \to \frac{1}{2|\vec{p}|} \delta(\omega - |\vec{p}|)$ so we are
left with
\begin{align}
	S_{++}[\Dhv](\omega) &= \frac{\kappa^2 \theta(\omega)}{8\pi \omega} \int_{-1}^1 \rd u 
	\left(
		\frac{\sin\left((1-u\right) \omega L/2)}{1-u}
  \right)^2 
	(1 - u^2)^2 \nonumber\\
							 &= \frac{\kappa^2 \theta(\omega)}{8\pi \omega} \int_{-1}^1 \rd u\, 
	\sin^2\bigg((1 - u)\frac{ \omega L}{2} \bigg) 
	(1 + u)^2
\end{align}
Using the integral
\begin{equation}
	\int_{-1}^1 \rd u\, \sin^2 \bigg( (1-u)\frac{x}{2} \bigg) (1+u)^2 =
	\frac{4}{3} - \frac{2}{x^2} + \frac{\sin 2x}{x^3}
\end{equation}
we get
\begin{equation}
	S_{++}[\Dhv](\omega) = \frac{\kappa^2 \theta(\omega)}{8\pi \omega} 
	\bigg( \frac{4}{3} - \frac{2}{(\omega L)^2} + \frac{\sin 2\omega L}{(\omega
	L)^3} \bigg) = S_{--}[\Dhv](\omega)
\end{equation}
For the $+-$ branch, the full integral to be
evaluated is similarly
\begin{equation}
	S_{+-}[\Dhv](\omega) = \frac{\kappa^2}{2}\pi e^{-i\omega L} \theta(\omega) 
				\int \frac{\rd^3 \vec{p}}{(2\pi)^3} 
        \left(
        \frac{ \cos (\vec{p}\cdot \vec{n}L) -  \cos (\omega L) }{(\omega^2  - (\vec{p}\cdot \vec{n})^2)}
        \right)
	\bigg( 1 - \frac{(\pn)^2}{\omega^2}\bigg)^2
	\delta(-\omega ^2 + |\vec{p}|^2)
\end{equation}
Using the same parametrization as above, it reduces to
\begin{align}
	S_{+-}[\Dhv](\omega)  &= \frac{\kappa^2 \theta(\omega)e^{-i\omega L}}{16\pi \omega} \int_{-1}^1
	\rd u\, \bigg( \frac{\cos(\omega L u) - \cos(\omega L)}{1-u^2} \bigg)
	(1-u^2)^2 \nonumber \\
								&= \frac{\kappa^2 \theta(\omega)e^{-i\omega L}}{16\pi \omega} \int_{-1}^1
	\rd u\, \bigg( \cos(\omega L u) - \cos(\omega L) \bigg)
	(1-u^2) \nonumber
\end{align}
Using the integral
\begin{equation}
	\int_{-1}^1 \rd u\, \bigg( \cos(x u) - \cos(x) \bigg) (1-u^2) 
	= 4 \bigg[ \frac{\sin x}{x^3} - \cos x \bigg( \frac{1}{3} + \frac{1}{x^2}\bigg) \bigg]
\end{equation}
we get
\begin{equation}
	S_{+-}[\Dhv](\omega)  = \frac{\kappa^2 \theta(\omega)e^{-i\omega L}}{4\pi \omega} 
	\bigg[ \frac{\sin (\omega L)}{(\omega L)^3} - \cos (\omega L) \bigg( \frac{1}{3} +
	\frac{1}{(\omega L)^2}\bigg) \bigg]
\end{equation}
Since $S_{-+} = S_{+-}^*$, their sum evaluates to
\begin{align}
	\big( S_{+-} + S_{-+}\big)[\Dhv](\omega)  &= \frac{\kappa^2 \theta(\omega)}{2\pi \omega} 
	\cos(\omega L)\bigg[ \frac{\sin (\omega L)}{(\omega L)^3} - \cos (\omega L) \bigg( \frac{1}{3} +
	\frac{1}{(\omega L)^2}\bigg) \bigg] \nonumber \\
    &= \frac{\kappa^2 \theta(\omega)}{4\pi \omega}
	 \bigg[ 
		 \frac{\sin (2\omega L)}{(\omega L)^3} - \cos(2\omega L) 
		 \bigg( 
			 \frac{1}{3} + \frac{1}{(\omega L)^2}
		 \bigg) 
		 - \frac{1}{(\omega L)^2} - \frac{1}{3}
	 \bigg]
\end{align}
Putting all the terms together, we obtain the spectrum from (\ref{eqn:SDh})
\begin{align}
	S[\Dhv](\omega) &= \bigg( 2S_{++} + S_{+-} + S_{-+}\bigg) [\Dhv](\omega) \\
		&= 
		\frac{\theta(\omega)\kappa^2}{4\pi\omega} 
		\bigg[
			1 - \frac{3}{(\omega L)^2} + \frac{2\sin(2\omega L)}{(\omega L)^3}
			- \cos(2\omega L)
			\bigg(\frac{1}{3} + \frac{1}{(\omega L)^2} \bigg) 
		\bigg] \nonumber
\end{align}

\subsection{Induced vacuum spectrum}
\label{app:B.3}

This appendix details the evaluation of the power spectrum for the induced
fluctuations in the vacuum state $S[\DHv]$ shown in (\ref{eqn:SDH}). The
computation follows the same steps as in appendix \ref{app:B.2}, with the
crucial difference that the fluctuation scalar \eqref{eqn:DHv}, which we restate
here as
\begin{equation}
	\DHv(\omega, \vec{p}) = \frac{\kappa^4}{5120 \pi}
	\theta(\omega) \theta(\omega^2 - |\vec{p}|^2)
	\bigg( 1- \frac{(\pn)^2}{\omega^2}\bigg)^2
\end{equation}
has support inside the future light cone due to $\theta(\omega^2 - |\vec{p}|^2)$
rather than on the light cone. This means that the integral with respect to
$|\vec{p}|$ is no longer a Dirac delta evaluation at $|\vec{p}| = \omega$, but
an integral in the region $|\vec{p}| \in [0, \omega]$. Using the same
parametrization as in appendix \ref{app:B.2}, we can write the full integral for
the $++$ branch as
\begin{align}
	S_{++}[\DHv](\omega) &= \frac{\kappa^4\theta(\omega)}{1280\pi} \int_{0}^{\omega}
	\frac{\rd|\vec{p}|\, |\vec{p}|^2}{(2\pi)^2} \int_{-1}^{1}\rd u\, 
	\frac{\sin^2[ ( 1 - \frac{|\vec{p}| u}{\omega} ) \frac{\omega L}{2} ]}
	{\omega^2\big(1 - \frac{|\vec{p}|u}{\omega} \big)^2}	
	\bigg( 1 - \bigg(\frac{|\vec{p}| u}{\omega} \bigg)^2\bigg)^2\\
											 &= \frac{\kappa^4\theta(\omega)}{1280\pi(2\pi)^2\omega^2}
											 \int_{0}^{\omega}
	\rd|\vec{p}|\, |\vec{p}|^2 \int_{-1}^{1}\rd u\, 
	\sin^2 \bigg[ \bigg( 1 - \frac{|\vec{p}| u}{\omega} \bigg) \frac{\omega L}{2} \bigg]
	\bigg( 1 + \frac{|\vec{p}| u}{\omega} \bigg)^2 \nonumber
\end{align}
Introducing $y := |\vec{p}|/\omega$, we can write
\begin{equation}
	S_{++}[\DHv](\omega) = \frac{\kappa^4\theta(\omega)\omega}{1280\pi(2\pi)^2}
	\int_0^1 \rd y\, y^2 \int_{-1}^1 \rd u\, 
	\sin^2 \bigg[ (1-y\, u) \frac{\omega L}{2}\bigg] (1+y\,u)^2
\end{equation}
Using the integral
\begin{equation}
	\int_0^1 \rd y\, y^2 \int_{-1}^1 \rd u\, \sin^2 
	\bigg[ (1-y\, u) \frac{x}{2}\bigg] (1+y\,u)^2 = \frac{2}{5} + \frac{2}{x^2} -
	\frac{9}{x^4} - \frac{3 \cos (2x)}{x^4} + \frac{6\sin (2x)}{x^5}
\end{equation}
we get
\begin{equation}
	S_{++}[\DHv](\omega) = \frac{\kappa^4 \theta(\omega)\omega}{5120\pi^3} 
	\bigg(
		\frac{2}{5} + \frac{2}{(\omega L)^2} -
		\frac{9}{(\omega L)^4} - \frac{3 \cos (2\omega L)}{(\omega L)^4} + \frac{6\sin
		(2\omega L)}{(\omega L)^5}
	\bigg)
\end{equation}
Similarly to the previous case in \ref{app:B.2}, $\DHv(\omega, -\vec{p}) =
\DHv(\omega, \vec{p})$ implies $S_{--}[\DHv] = S_{++}[\DHv]$. For the $+-$ branch,
the full integral is
\begin{align}
	S_{+-}[\DHv](\omega) &= \frac{\kappa^4 \theta(\omega)e^{-i\omega L}}{2560 \pi} 
	\int_{0}^{\omega} \frac{\rd|\vec{p}|\, |\vec{p}|^2}{(2\pi)^2} \int_{-1}^{1}\rd u\,
	\bigg( \frac{\cos(|\vec{p}| u L) - \cos(\omega L)}{\omega^2 - (|\vec{p}|
	u)^2}\bigg) \bigg( 1 - \bigg( \frac{|\vec{p}| u}{\omega}\bigg)^2\bigg)^2
	\nonumber\\
											 &= \frac{\kappa^4 \theta(\omega)e^{-i\omega L}}{2560 \pi(2\pi)^2\omega^2} 
	\int_{0}^{\omega} \rd|\vec{p}|\, |\vec{p}|^2 \int_{-1}^{1}\rd u\,
	\bigg( \cos(|\vec{p}| u L) - \cos(\omega L)\bigg) \bigg( 1 - \bigg(
	\frac{|\vec{p}| u}{\omega}\bigg)^2\bigg)
\end{align}
After the same change of variables, we get
\begin{equation}
	S_{+-}[\DHv](\omega) = \frac{\kappa^4 \theta(\omega)e^{-i\omega L}\omega}{2560 \pi(2\pi)^2} 
	\int_{0}^{1} \rd y\,y^2 \int_{-1}^{1}\rd u\,
	\bigg( \cos(y\, u\, \omega L) - \cos(\omega L)\bigg) (1 - (y\, u)^2)
\end{equation}
Using the integral
\begin{equation}
	\int_{0}^{1} \rd y\,y^2 \int_{-1}^{1}\rd u\,
	\bigg( \cos(y\, u\, x) - \cos(x)\bigg) (1 - (y\, u)^2) =
	8 \bigg[ \sin x\bigg( \frac{3}{x^5} - \frac{1}{x^3}\bigg) - \cos x
	\bigg( \frac{1}{15} + \frac{3}{x^4}\bigg)\bigg]
\end{equation}
so we get
\begin{equation}
	S_{+-}[\DHv](\omega) = \frac{\kappa^4 \theta(\omega)e^{-i\omega L}\omega}{1280
	\pi^3}
	\bigg[ \sin (\omega L)\bigg( \frac{3}{(\omega L)^5} - \frac{1}{(\omega L)^3}\bigg) - \cos (\omega L)
	\bigg( \frac{1}{15} + \frac{3}{(\omega L)^4}\bigg)\bigg]
\end{equation}
and then together with $S_{-+}$
\begin{align}
	\big( S_{+-} + S_{-+} \big) [\DHv](\omega) &= \frac{\kappa^4
	\theta(\omega)\omega\, \cdot 2\cos(\omega L)}{1280
	\pi^3}
	\bigg[ \sin (\omega L)\bigg( \frac{3}{(\omega L)^5} - \frac{1}{(\omega L)^3}\bigg) - \cos (\omega L)
	\bigg( \frac{1}{15} + \frac{3}{(\omega L)^4}\bigg)\bigg] \nonumber \\
																						 &= \frac{\kappa^4
	\theta(\omega)\omega}{1280\pi^3}
	\bigg[ -\frac{1}{15} - \frac{3}{(\omega L)^4} 
		+ \sin (2\omega L)\bigg( \frac{3}{(\omega L)^5} - \frac{1}{(\omega L)^3}\bigg)\nonumber\\
																						 & - \cos (2\omega L)
	\bigg( \frac{1}{15} + \frac{3}{(\omega L)^4}\bigg)\bigg] 
\end{align}
Finally, adding all the contributions together, we obtain (\ref{eqn:SDH})
\begin{align}
	S[\DHv](\omega) &= \bigg( 2 S_{++} + S_{+-} + S_{-+} \bigg)[\DHv](\omega) \\
	&= \frac{\kappa^4
	\theta(\omega)\omega}{1280\pi^3}
	\bigg[ \frac{2}{15} +\frac{1}{(\omega L)^2} - \frac{15}{2(\omega L)^4} -
	\cos(2\omega L)\bigg(\frac{1}{15} + \frac{9}{2(\omega L)^4}\bigg)  +
	\sin(2\omega L)\bigg(
\frac{6}{(\omega L)^5}- \frac{1}{(\omega L)^3}\bigg)  \bigg] \nonumber
\end{align}

\subsection{Infinite-time limit and non-translation-invariant contributions}
\label{app:B:wienerkhinchin}

In this appendix, we show how the operational definition of the spectral density
as an infinite-time limit of windowed measurements \eqref{eqn:Sdef} reduces to the
Wiener--Khinchin form for stationary processes \eqref{eqn:Stau}. We then apply the same
manipulations to autocorrelation functions that depend on $t+t'$ (rather than
$t-t'$), which arise from the non-translation-invariant (NTI) contributions in
the squeezed state, and show that these terms do not contribute to the spectrum
at nonzero frequencies.

Let $x(t)$ be a stochastic process measured for a finite time $T$ with a window
function $w_T$. The windowed Fourier transform is
\begin{equation}
	\tilde{x}_T(\omega) := \int_{-\infty}^{\infty} \rd t\,
	w_T(t)\,e^{-i\omega t}\,x(t) ,
\end{equation}
and the finite-time spectrum is
\begin{equation}
	S_T(\omega) := \frac{1}{\|w_T\|^2}
	\Big\langle \tilde{x}_T(\omega)\tilde{x}_T^\dagger(\omega)\Big\rangle,
	\qquad
\end{equation}
Expanding the Fourier transforms gives
\begin{equation}
	S_T(\omega)=\frac{1}{\|w_T\|^2}
	\int_{-\infty}^{\infty}\rd t
	\int_{-\infty}^{\infty}\rd t'\,
	e^{-i\omega(t-t')}\,
	w_T(t)w_T(t')\,W(t,t'),
	\label{eqn:ST_W_def}
\end{equation}
where $W(t,t'):=\langle x(t)x(t')\rangle$.

\subsubsection{Stationary case}
Assume stationarity, so $W(t,t')=W(t-t')$. Introduce the Fourier representation
\begin{equation}
	W(t-t')=\int_{-\infty}^{\infty}\frac{\rd\Omega}{2\pi}\,
	e^{i\Omega(t-t')}\,f(\Omega),
	\label{eqn:W_stationary_f}
\end{equation}
where $f(\Omega)$ is the spectral distribution associated to the stationary
autocorrelation. Substituting this into
\eqref{eqn:ST_W_def} and performing the $t,t'$ integrals yields
\begin{align}
	S_T(\omega)
	&=\frac{1}{\|w_T\|^2}
	\int\frac{\rd\Omega}{2\pi}\,f(\Omega)
	\left(\int_{-\infty}^{\infty}\rd t\,w_T(t)\,
	e^{it(\Omega-\omega)}\right)
	\left(\int_{-\infty}^{\infty}\rd t'\,w_T(t')\,
	e^{-it'(\Omega-\omega)}\right) \nonumber \\
	&=\int\frac{\rd\Omega}{2\pi}\,f(\Omega)\,
	\frac{|\tilde w_T(\Omega-\omega)|^2}{\|w_T\|^2}.
	\label{eqn:ST_stationary_kernel}
\end{align}
Thus, the stationary spectrum is controlled by the kernel
\begin{equation}
	J_T(\Omega):=\frac{|\tilde w_T(\Omega)|^2}{\|w_T\|^2}.
	\label{eqn:JT_def}
\end{equation}
which limits to a delta distribution with support at $\Omega=0$.
To understand the infinite-time limit, we assume that $w_T(t)=w(t/T)$. Then
\begin{equation}
	\tilde w_T(\nu)=T\,\tilde w(T\nu),
	\qquad
	\|w_T\|^2=T\,\|w\|^2.
	\label{eqn:scaling_w}
\end{equation}
Using this, the kernel becomes
\begin{equation}
	J_T(\Omega)=\frac{T}{\|w\|^2}\,|\tilde w(T\Omega)|^2.
	\label{eqn:JT_scaled}
\end{equation}
and then
\begin{align}
	S_T(\omega) &= \int_{-\infty}^{\infty}
	\frac{\rd\Omega}{2\pi} \,f(\Omega)\,J_T(\Omega-\omega) \\
	&=\frac{T}{\|w\|^2}\int_{-\infty}^{\infty} \frac{\rd\Omega}{2\pi}\,
	f(\Omega)\, |\tilde w(T(\omega-\Omega))|^2 \nonumber\\
	&=\frac{1}{\|w\|^2}\int_{-\infty}^{\infty} \frac{\rd u}{2\pi}\,
	f\Big(\omega-\frac{u}{T}\Big)\,|\tilde w(u)|^2 \nonumber
	\label{eqn:JT_delta_limit}
\end{align}
where we changed variables $u:=T(\omega-\Omega)$ in the last step. The limit $T
\to \infty$ can be taken pointwise in the integrand as long as the conditions of
dominated convergence are satisfied, which requires that there exists an
integrable function $g(u) \in L^1(\mathbb{R})$ such that $g(u) \geq \big|
f\Big(\omega-\frac{u}{T}\Big)\big|\,|\tilde w(u)|^2 $ \cite{Folland1984RealAM}, with the bound being uniform in $T$ for $T>1$. 
 If $f$ has a maximum,
which is the case in all of section \ref{sec:3} since $F_h$ from \eqref{eqn:Fh}
has a maximum, one can choose $g(u) = f_\text{max}|\tilde w(u)|^2$ for any
window function since the integrability condition becomes the same as the existence
of the norm. 
Under these conditions, the limit may be taken inside the integral, yielding
\begin{align}
	S(\omega) &= \frac{1}{\|w\|^2}\int_{-\infty}^{\infty} \frac{\rd u}{2\pi}\,
	\lim_{T \to \infty }f\Big(\omega-\frac{u}{T}\Big)\,|\tilde w(u)|^2 \nonumber \\
						&= f(\omega)\frac{1}{\|w\|^2}\int_{-\infty}^{\infty} \frac{\rd u}{2\pi}\,
	|\tilde w(u)|^2 \nonumber \\
						&= f(\omega)
\end{align}
where we used Parseval's theorem $\int \rd u\,|\tilde w(u)|^2 = 2\pi\int \rd
t\,|w(t)|^2 = 2\pi\,\|w\|^2$.
The equality $S(\omega) = f(\omega)$ is precisely the Wiener--Khinchin
identification of the spectral density with the Fourier transform of the
stationary autocorrelation function as used in \eqref{eqn:Stau} since in our
setup $W(t-t') = \langle \delta \tau(t) \delta \tau(t')\rangle$ and thus its
Fourier representation $f(\Omega)$ is what we compute with the formula
\eqref{eqn:SG}.

The situation is more delicate for the  spectrum of induced fluctuations described by
$F_H$ from \eqref{eqn:FH}. The spectrum is linearly
increasing as $F_H(x)\sim \theta(x) x$, and a bound looks like $|F_H(\omega -
\frac{u}{T})| \leq C(1 + |u|)$ for $T>1$ and some positive constant $C$. 
The condition of dominated convergence can only be satisfied if we choose window functions that decay rapidly enough at large $u$. A sufficient condition is that the moment $\int \rd u |u| \, |\tilde{w}(u)|^2$ exists.
Due to the increase in $F_H$, a box function $w(t) = \theta(1-|t|)$ is not
sufficient in this example as it only decays as $u^{-1}$. A triangular window
$w(t) = \theta(1-|t|)(1-|t|)$ works as it has $u^{-2}$ decay.

\subsubsection{Non-translation-invariant contributions}
We now consider the case relevant for squeezed states, where the NTI part of the
autocorrelation depends on $t+t'$:
\begin{equation}
	W_{\text{NTI}}(t,t') = W_{\text{NTI}}(t+t')
	=\int_{-\infty}^{\infty}\frac{\rd\Omega}{2\pi}\,e^{i\Omega(t+t')}\,f(\Omega).
	\label{eqn:W_NTI_f}
\end{equation}
In our case, this form arises from the NTI contribution to the fluctuation
scalar from \eqref{eqn:WNTI}
\begin{equation}
    \DNTI(p) = \pi \kappa^2 \Sigma_n^2(p)\big[ \delta_+(p^2) M(p) + \delta_-(p^2) M^*(p)\big],
\end{equation}
together with the inverse Fourier representation of the time delay, which yields
\begin{equation}
	f^\text{NTI}(\Omega) := \int \frac{\rd^3 \vec{p}}{(2\pi)^3}\,
	F(\Omega,\vec{p})F(\Omega,-\vec{p})\,\DNTI(\Omega,\vec{p}).
\end{equation}
This is the analogue of \eqref{eqn:SG} corresponding to the time-dependence
$t+t'$. Substituting \eqref{eqn:W_NTI_f} into \eqref{eqn:ST_W_def} produces
\begin{align}
	S_T^{\text{NTI}}(\omega)
	&=\frac{1}{\|w_T\|^2}
	\int\frac{\rd\Omega}{2\pi}\,f^\text{NTI}(\Omega)
	\left(\int_{-\infty}^{\infty}\rd t\,w_T(t)\,
	e^{-it(\omega-\Omega)}\right)
	\left(\int_{-\infty}^{\infty}\rd t'\,w_T(t')\,
	e^{-it'(\omega+\Omega)}\right) \nonumber\\
	&=\int\frac{\rd\Omega}{2\pi}\,f^\text{NTI}(\Omega)\,
	\frac{\tilde w_T(\omega-\Omega)\tilde w_T^*(\omega+\Omega)}{\|w_T\|^2}.
	\label{eqn:ST_NTI_kernel}
\end{align}
where we now have the NTI kernel
\begin{equation}
	K_T(\Omega):=
	\frac{\tilde w_T(\omega-\Omega)\tilde w_T^*(\omega+\Omega)}{\|w_T\|^2}.
	\label{eqn:KT_def}
\end{equation}
Using the same scaling assumption \eqref{eqn:scaling_w}, we may write
\begin{equation}
	K_T(\Omega)
	=\frac{T}{\|w\|^2}\,
	\tilde w\big(T(\omega-\Omega)\big)\,
	\tilde w^*\big(T(\omega+\Omega)\big).
	\label{eqn:KT_scaled}
\end{equation}
For fixed $\omega\neq 0$, the first factor localizes around $\Omega=\omega$,
while the second localizes around $\Omega=-\omega$. Since these localization
points are separated, their product has vanishing overlap in the $T\to\infty$
limit. More precisely, with the same change of variables $u:=T(\omega-\Omega)$,
we write
\begin{equation}
	S_T^\text{NTI}(\omega) =
	\frac{1}{\|w\|^2}\int_{-\infty}^{\infty} \frac{\rd u}{2\pi}\,
	f^\text{NTI}\Big(\omega-\frac{u}{T}\Big)\,
	\tilde w(u)\,
	\tilde w^*\!\big(2T\omega-u\big)
	\label{eqn:KT_pairing}
\end{equation}
Upon taking the limit, we may again use dominated convergence provided there
exists an integrable function $g(u) \in L^1(\mathbb{R})$ such that $g(u) \geq
\big| f^\text{NTI}\Big(\omega-\frac{u}{T}\Big)\,\tilde w(u)\tilde
w^*(2T\omega - u)\big|$. Since $\DNTI$ is light cone supported, $f^\text{NTI}$
will have characteristics similar to $F_h$ from \eqref{eqn:Fh} and the same
argument works as above. We can therefore write
\begin{equation}
	S^\text{NTI}(\omega) = 
	\frac{1}{\|w\|^2}\int_{-\infty}^{\infty} \frac{\rd u}{2\pi}\,
	\lim_{T \to \infty} f^\text{NTI}\Big(\omega-\frac{u}{T}\Big)\,
	\tilde w(u)\,
	\tilde w^*\!\big(2T\omega-u\big) 
\end{equation}
The property of window functions is that their Fourier transform is sharply
peaked as their argument tends to zero and decay away from it so we have
\begin{equation}
	\lim_{T \to \infty} f^\text{NTI}\Big(\omega-\frac{u}{T}\Big)\tilde w(u)\,
	\tilde w^*\!\big(2T\omega-u\big) = 0
\end{equation}
as long as $\omega \neq 0$. Therefore, the integral \eqref{eqn:KT_pairing}
vanishes as $T\to\infty$, and we conclude that
\begin{equation}
	S^{\text{NTI}}(\omega):=\lim_{T\to\infty}S_T^{\text{NTI}}(\omega)=0, \quad
	\omega \neq 0
\end{equation}
with at most a residual contribution localized at $\omega=0$.

This establishes that the non-translation-invariant ($t+t'$) contributions do
not affect the power spectral density at nonzero frequency, while the stationary
($t-t'$) component reduces to the standard Wiener--Khinchin spectrum.

%% file: appendices/appC_correlators.tex
\section{Correlation function of induced fluctuations}
\label{app:C}

In this appendix, we detail derivations of the induced correlation functions. First, we evaluate the Fourier transform of the two-point function of the convolution, and then provide detailed calculations of the resulting integrals.

\subsection{Fourier transform}
\label{app:C.1}

Our goals is to evaluate the Fourier transform of the
correlation function $\Gv$ as defined in (\ref{eqn:Gv})
\begin{equation}
	\Gv = \frac{1}{4}\int \rd^4 \bar{x}\, e^{ip\bar{x}}
        \int \rd^4 y \int\rd^4 y' G_R(x-y)G_R(x'-y') 
        \langle \bar{T}_{ab}(y)\bar{T}_{cd}(y')\rangle
\end{equation}
which amounts to
evaluating the convolution with the retarded Green's function and the Fourier
transform. First, we use that the normal ordered stress tensor takes the form
\begin{align}
    :\bar{T}_{ab}(y): &= 
        \int \measure{p}\measure{q} \\
        &
        \times p_a q_b 
        \bigg(
            a_q^\dagger a_p \, e^{-i(p-q)y} 
            + a^\dagger_p a_q e^{i(p-q)y}  
            - a_p a_q e^{-i(p+q)y} 
            - a^\dagger_q a^\dagger_p e^{i(p+q)y}
        \bigg) \nonumber
\end{align}
Multiplying by $:\bar{T}_{cd}(y'):$ and taking the vacuum expectation value gives
\begin{equation}
    \langle \bar{T}_{ab}(y)\bar{T}_{cd}(y')\rangle
        = \partial_a \partial_c  W^{(0)}(y-y') 
          \partial_b \partial_d W^{(0)}(y-y') 
        + (c \leftrightarrow d)
\end{equation}
with $W^{(0)}(y-y')$ being the vacuum Wightman function of the scalar field 
\begin{equation}
    W^{(0)}(x-x') = \int\frac{\rd^4 p}{(2\pi)^3}\delta_+(p^2)e^{-ip(x-x')}
\end{equation}
We also need the Fourier transform of the retarded propagator given by
\begin{equation}
   G_{R}(x-y) = \int\frac{\rd^4 p}{(2\pi)^4} \frac{e^{-ip(x-y)}}{-(p^0 + i\epsilon)^2 + \vec{p}^2} 
\end{equation}
Let us introduce the Fourier transform $\tilde{G}_R^{-1}(p) := -(p^0 + i \epsilon)^2 + \vec{p}^2$,
and we also have
\begin{equation}
    \partial_a \partial_c W^{(0)}(y-y') = 
        \int \frac{\rd^4 p}{(2\pi)^3}\, \delta_+(p^2) (-p_a p_c)\, e^{-ip(y-y')}
\end{equation}
We then introduce $p_{1,2}$ as the conjugate momenta to $x-y$ and $x'-y$ in the
Green's functions and $p_{3,4}$ as the momenta in the Wightman function so that
we can write
\begin{align}
    \Gv
    &= 
		\frac{1}{4}\int \rd^4 \bar{x} \, e^{ip \bar{x}} 
    \int \rd^4 y \int \rd^4 y' 
    \int \frac{\rd^4 p_1}{(2\pi)^4} 
        \tilde{G}_R(p_1)\, e^{-ip_1(x-y)}
    \int \frac{\rd^4 p_2}{(2\pi)^4} 
        \tilde{G}_R(p_2)\, e^{-ip_2(x'-y)} \nonumber \\
    &\times 
    \int \frac{\rd^4 p_3}{(2\pi)^3}
    \int \frac{\rd^4 p_4}{(2\pi)^3}\, e^{-i(p_3+p_4)(y-y')}
    \delta_+(p_3^2)\delta_+(p_4^2) 
    \bigg( (p_3)_a (p_3)_c(p_4)_b (p_4)_d + (c\leftrightarrow d)\bigg)
\end{align}
Evaluating the $y$ and $y'$ integrals gives the momentum conservations 
$p_1= -p_2= p_3+p_4$.
So that after evaluating the $\int \rd^4 p_2$ integral one has 
\begin{align}
	\Gv  &= \frac{1}{4}
    \int \rd^4 \bar{x} \,  
    \int \rd^4 p_1 \,
    \tilde{G}_R(p_1)\tilde{G}_R(-p_1)\, e^{-i(p_1-p) \bar{x}} \\
    &\times
    \int \frac{\rd^4 p_3}{(2\pi)^3}
    \int \frac{\rd^4 p_4}{(2\pi)^3}\,  \delta^{(4)}(p_1 - p_3 - p_4)\, 
    \delta_+(p_3^2)\delta_+(p_4^2)  
    \bigg( (p_3)_a (p_3)_c(p_4)_b (p_4)_d + (c \leftrightarrow d)\bigg) \nonumber
\end{align}
Now we evaluate the $\int \rd^4\bar{x}$ integral which identify $p$ and $p_1$.
We are left with the  integrals over $p_3,p_4$ so that
\begin{align}
    \Gv &=
		\frac{\tilde{G}_R(p) \tilde{G}_R(-p)}{4}
    \int \rd^4 p_3 \int \frac{\rd^4 p_4}{(2\pi)^2}\, 
    \delta^{(4)}(p - p_3 - p_4)\, 
    \delta_+(p_3^2)\delta_+(p_4^2)  \nonumber \\ 
    &\times \bigg( (p_3)_a (p_3)_c(p_4)_b (p_4)_d + (c \leftrightarrow d)\bigg) \nonumber \\
    &= \frac{\tilde{G}_R(p) \tilde{G}_R(-p)}{4(2\pi)^2}
        \int \rd^4 k \, 
        \delta_+[(p-k)^2]\, \delta_+(k^2)\, 
        \bigg( (p-k)_a(p-k)_c k_b k_d  + (c \leftrightarrow d) \bigg) \nonumber\\
    &= \frac{\tilde{G}_R(p) \tilde{G}_R(-p)}{4(2\pi)^2} \int \rd \Phi_p(k) 
        \bigg(s_a s_c k_b k_d + (c \leftrightarrow d) \bigg)\bigg|_{s = p-k}
\end{align}
where we introduced
\begin{equation}
    \rd\Phi_p(k) := \rd^4 k\, \delta_+(s^2)\, \delta_+(k^2)\bigg|_{s=p-k}
\end{equation}
Looking at the product of the retarded propagators
\begin{align}
    \tilde{G}_R(p) \tilde{G}_R(-p) &= \frac{1}{
        \big( -(p^0  + i \epsilon)^2 + \vec{p}^2 \big)
        \big( -(-p^0 + i \epsilon)^2 + \vec{p}^2 \big) } \nonumber \\
        &= \frac{1}{(p^2)^2 + 2\epsilon^2( \big(p^0)^2 + \vec{p}^2 \big) + \cO(\epsilon^4)}
\end{align}
we can just take the $\epsilon \to 0$ limit so that 
\begin{equation}
	\lim_{\epsilon \to 0}\tilde{G}_R(p) \tilde{G}_R(-p)  = \frac{1}{(p^2)^2}
\end{equation}
Note that if we chose advanced propagation instead with $\tilde{G}_A^{-1}(p) := -(p^0 - i \epsilon)^2 + \vec{p}^2$, we would get the same result as $\tilde{G}_R(p) \tilde{G}_R(-p) = \tilde{G}_A(p) \tilde{G}_A(-p)$. The main integral we have left to calculate is thus
\begin{equation}
    \int \rd \Phi_p(k) 
        \bigg(s_a s_c k_b k_d + (c \leftrightarrow d) \bigg)\bigg|_{s = p-k}
\end{equation}
which are called the two-particle phase space integrals \cite{Schwartz:2014sze}
and we detail their calculations in the next subsection.

\subsection{Two-particle phase space integral}
\label{app:C.2}

In this subsection, we detail the evaluation of the integral
\begin{equation}
    \int \rd \Phi_p(k) 
        \bigg(s_a s_c k_b k_d + (c \leftrightarrow d) \bigg)\bigg|_{s = p-k}
\end{equation}
First, we consider the simplest scalar case
\begin{equation}
	I(p) := \int \rd \Phi_p(k)
\end{equation}
A simple way to calculate this integral is by using Lorentz invariance
\begin{equation}
    I(\Lambda p) = I(p), \quad \forall \Lambda \in O^+(1,3)
\end{equation}
which means that the integral can be evaluated on a representative of each causal class for $p$. For concreteness we  choose  $p_1=\alpha(1,\vec{0}),$ for a timelike representative and $ p_2 =
\beta(0, \hat{e}_x)$ with $p_1^2 = -\alpha^2,\, p_2^2 = \beta^2$, for a spacelike one, where
$\hat{e}_x$ is a unit 3-vector in the $x$ direction.  For a general $p$, we
write
\begin{align}
   I(p) 
   &= \int \rd k^0 \int \rd^3 \vec{k}\, 
    \theta(k^0)\frac{1}{2|\vec{k}|}\delta(k^0 - |\vec{k}|)
    \theta(p^0-k^0)\delta(p^2 + k^2 - 2p\cdot k) \\
   &= \int_{|\vec{k}|< p^0} \rd^3 \vec{k}\, 
    \frac{1}{2|\vec{k}|} \delta(p^2 + 2p^0 |\vec{k}| - 2 \vec{p}\cdot \vec{k}) \nonumber
\end{align}
For a timelike $p$ we see that the integral vanishes if $p$ is not future pointing. 
Substituting $p_1$
\begin{align}
   I(p_1) 
   &= \theta(\alpha) \int_{|\vec{k}|< \alpha} \rd^3 \vec{k}\, 
    \frac{1}{2|\vec{k}|}  \delta(-\alpha^2 + 2\alpha |\vec{k}|) \\
   &= 4\pi \theta(\alpha) \int_{|\vec{k}|< \alpha} \rd |\vec{k}|\, 
        \frac{|\vec{k}|}{2}  
        \frac{1}{2 \alpha} \delta\bigg(|\vec{k}|- \frac{\alpha}{2} \bigg) \nonumber \\
   &= \frac{\pi}{2} \theta(\alpha) \nonumber \\
   &= \frac{\pi}{2} \theta_-(-p_1^2) \nonumber
\end{align}
 For a spacelike $p$ we substitute $p_2$, we get
\begin{equation}
    I(p_2) 
    = \int \rd^3 \vec{k}\, 
    \frac{1}{2|\vec{k}|} \theta(-|\vec{k}|) \delta(\beta^2 - 2 \beta \hat{e}_x \cdot \vec{k})  
    = 0
\end{equation}
Therefore, the scalar piece has a constant support on future-directed timelike
vectors
\begin{equation}
	I(p) = \frac{\pi}{2}\theta_+(p)
\end{equation}
where $\theta_+(p) := \theta(p^0)\theta(-p^2)$ is the projection onto future-directed
timelike vectors. 
The case where $p$ is null is  of measure $0$ so it doesn't affect the integral.

To generalize $I(p)$, we now use a covariant parametrization to evaluate
\begin{equation}
	I^{(F)}{}_{abc...}(p) := \int \rd \Phi_p(k)\, F_{abc...}(k,s)
\end{equation}
This integral vanishes if $p$ is not future timelike.
To see this we use that the solution of the equations $s^2=k^2=0$ and $s+k=p$ are given by 
\be 
s= \alpha_p (u+n), \qquad 
k= \alpha_p (u-n),
\ee 
where $u,n$ are normalized orthogonal vectors $u^2=-\epsilon$, $n^2=\epsilon$, with $\epsilon=\pm1$ and $u\cdot n=0$; while $u$ is parallel to $p$ with $p= 2\alpha_p u$. The causality condition that both $s$ and $k$ are future pointing means that $\epsilon=+1$ with $u$ future pointing and $\alpha_p>0$.

To evaluate the measure in the case where $p$ is future timelike, we decompose
\begin{equation}
    k = \lambda\, u + \mu \, n
\end{equation}
with $u^a := \frac{p^a}{2\alpha_p}$, $\alpha_p:=\frac12 \sqrt{-p^2}$, $u^2 = -1$, $n^2 = 1$, $u\cdot n = 0$ and
$\lambda,\mu$ are scalar functions. With this decomposition, we can write the
norms of $k$ and $s=p-k$ as
\begin{equation}
    k^2 = 
        (\mu - \lambda)(\mu + \lambda), 
    \quad 
    s^2 = 
        \bigg( \mu - (2\alpha_p- \lambda) \bigg) 
        \bigg( \mu + (2\alpha_p - \lambda) \bigg)
\end{equation}
The $\theta$ functions in $\delta_+(k^2),\, \delta_+(s^2)$ restrict only to one
of the branches $\lambda, \mu >0$, therefore giving
\begin{subequations}
	\begin{align}
		 \delta_+(k^2) &= \theta(-k\cdot t) \frac{1}{2|\mu|} 
					\delta(\lambda - \mu) \\
		 \delta_+(s^2) &= \theta(-s \cdot t) \frac{1}{2|\mu|} 
					\delta \bigg( (2\alpha_p- \lambda) - \mu \bigg)
	\end{align}
\end{subequations}
We also have the decomposition of the measure
\begin{equation}
    \rd^4 k = \rd \lambda\, \mu^2 \rd \mu \, \rd^2 \hat{n}
\end{equation}
This parametrization reproduces our previous result for $I(p)$:
\begin{align}
    I(p) &= \int \rd \lambda\, \mu^2 \rd \mu \, \rd^2 \hat{n}
        \theta(-k\cdot t) \frac{1}{2|\mu|} 
                \delta(\lambda - \mu)
        \theta(-s \cdot t) \frac{1}{2|\mu|} 
                \delta \bigg( (2\alpha_p- \lambda) - \mu \bigg) \nonumber \\
    &= \frac{1}{8} \int \rd^2 \hat{n} \, \theta(-k\cdot t) \theta(-s\cdot t) 
        \bigg|_{k = \alpha_p(u + n),\, s=\alpha_p(u - n)} \nonumber \\
    &= \frac{\pi}{2} \theta(-p\cdot t) = \frac{\pi}{2} \theta_-(-p^2).
\end{align}
To write the final form, we used that our
parametrization in terms of $u$ assumed that $p$ is timelike. This easily
generalizes to the insertion of a generic Lorentz covariant tensor
$F_{abc...}(k,s)$ as the final $\int \rd^2 \hat{n}$ integral is the average on
the unit 2-sphere with the arguments $k,s$ of $F$ are determined in terms of
$\alpha_p,u,n$:
\begin{equation}
    I^{(F)}{}_{abc...}(p) 
    = I(p) 
        \langle F_{abc...}\rangle_n
\end{equation}
where the notaton for the average
\begin{equation}
	\langle F_{abc...}\rangle_n := \langle F_{abc...}(k=\alpha_p(u + n), s = \alpha_p(u -
	n))\rangle_n = \frac{\int \rd^2 \hat{n}\,F_{abc...}(\hat{n})}{4\pi}
\end{equation}
compute the relevant averages, we need the following identities
\begin{subequations}
    \begin{align}
        \langle n_a \rangle &= 0 \\
        \langle n_a n_b \rangle &= \frac{1}{3} P_{ab} \\
        \langle n_a n_b n_c \rangle &= 0 \\
        \langle n_a n_b n_c n_d \rangle 
        &= \frac{1}{15}(P_{ab} P_{cd} 
        + P_{ac} P_{bd} 
        + P_{ad} P_{bc})
    \end{align}
\end{subequations}
with $P_{ab} := \eta_{ab} + u_a u_b$.
These come from the lift of the 3-dimensional results under the transverse projectors $e^i{}_a e^j{}_b \delta_{ij} = P_{ab}$ 

\begin{subequations}
    \begin{align}
        \langle n_i \rangle &= 0 \\
        \langle n_i n_j \rangle &= \frac{1}{3}\delta_{ij} \\
        \langle n_i n_j n_k \rangle &= 0 \\
        \langle n_i n_j n_k n_l \rangle &= \frac{1}{15}(\delta_{ij}\delta_{kl} + \delta_{ik}\delta_{jl} + \delta_{il}\delta_{jk})
    \end{align}
\end{subequations}
Specifically, we need the case when 
\begin{equation}
    F_{abcd}(k,s) = s_a s_c k_b k_d + (c \leftrightarrow d) 
\end{equation}
Under the average, $s_a k_b$ evaluates to
\begin{equation}
	s_a k_b = \alpha_p^2(u-n)_a (u+n)_b = \alpha_p^2\big( \Delta_{ab} - P_{ab} + u_a u_b + u_a n_b -
	u_b n_a \big)
\end{equation}
with $\Delta_{ab} := P_{ab} - n_a n_b = \eta_{ab} + u_a u_b - n_a n_b$. 
Since only even number of $n$'s have non-vanishing average, upon writing the
expression $s_a s_c k_b k_d $, only the following terms survive in the average
\begin{align}
	\alpha_p^{-4} s_a s_c k_b k_d &= \Delta_{ab} \Delta_{cd} + P_{ab}P_{cd} - \Delta_{ab}P_{cd}-
	\Delta_{cd} P_{ab} + u_a u_b u_c u_c +
	(\Delta-P)_{ab}u_c u_d + (\Delta-P)_{cd} u_a u_b
	\nonumber \\
									&+ u_a u_c (P-\Delta)_{bd} + u_b u_d (P-\Delta)_{ac} - u_a u_d
									(P-\Delta)_{bc} - u_c u_b (P-\Delta)_{ad}
\end{align}
Upon adding the contribution with $c,d$ exchanged, all the terms in the second
line above cancel. Using the averages
\begin{subequations}
\begin{align}
    \langle \Delta_{ab}\rangle &= \frac{2}{3}P_{ab} \\
    \langle (P-\Delta)_{ab}\rangle &= \frac{1}{3}P_{ab} \\
    \langle \Delta_{ab}\Delta_{cd}\rangle &= \frac{2}{5} P_{ab} P_{cd} + \frac{1}{15}(P_{ac}P_{bd}+ P_{ad}P_{bc}) 
\end{align}
\end{subequations}
we arrive at

\begin{equation}
	\big\langle s_a s_c k_b k_d + (c \leftrightarrow d) \big\rangle_n = \alpha_p^4
	\bigg[ 2 u_a u_b u_c u_d -
	\frac{2}{3} 
	\bigg( 
		P_{ab} u_c u_d + P_{cd} u_a u_b
	\bigg) + \frac{2}{15}
	\bigg(
    P_{ab}P_{cd}+ P_{ac}P_{bd} + P_{ad} P_{bc} 
\bigg)\bigg]
\end{equation}
so then our final integral evaluates to
\begin{align}
    \int \rd \Phi_p(k) 
		\bigg(s_a s_c k_b k_d + (c \leftrightarrow d) \bigg) &= 2\alpha_p^4 I(p) 
				\bigg[
				 u_a u_b u_c u_d -
	\frac{1}{3} 
	\bigg( 
		P_{ab} u_c u_d + P_{cd} u_a u_b
	\bigg) \nonumber \\
	&+ \frac{1}{15}
	\bigg(
    P_{ab}P_{cd}+ P_{ac}P_{bd} + P_{ad} P_{bc} 
 \bigg)
	\bigg]
\end{align}
where $\alpha_p^4 = (p^2)^2/16$.

%% file: appendices/appD_wightman.tex
\section{Wightman functions calculations}
\label{app:D}

In this appendix, we detail the derivations of the Wightman function for the
thermal and squeezed states of the gravitational field.

\subsection{Thermal Wightman function}
\label{app:D:thermal}

We compute the Wightman function for thermal state (\ref{eqn:thermalWightman}).
The difference between a free massless scalar field and the graviton computation
is only in the appearance of the polarization sum $\Pi_{abcd}$ from
(\ref{eqn:polsum}), and therefore we only present the computation for the scalar
field for convenience and then extend it to the gravitational case. The thermal
Wightman function is defined as 
\begin{equation}
    W^{(\beta)}(x-x') := \Tr \bigg( \rho_\beta \, \varphi(x)\varphi(x')\bigg)
\end{equation}
with $\rho_\beta = \frac{1}{Z} e^{-\beta H}$ and the Hamiltonian 
\begin{equation}
    H = \int \frac{\rd^3 \vec{p}}{(2\pi)^3} E_p\, a_p^\dagger a_p
\end{equation}
The mode expansion is
\begin{equation}
    \varphi(x) = \int \frac{\rd^3 \vec{p}}{(2\pi)^3\sqrt{2E_p}}\bigg(a_pe^{-ipx} + a_p^\dagger e^{ipx} \bigg)\bigg|_{p^0=E_\vec{p}},\quad E_\vec{p} = |\vec{p}|
\end{equation}
The partition function is calculated in the Fock basis in a box of volume $V$ as an IR regulator
\begin{equation}
        Z = \Tr e^{-\beta H} = \sum_{\{n_k \}} \bra{n_k\}} e^{-\beta \sum_\vec{p} E_p n_p} \ket{\{ n_k\}} 
        = \prod_\vec{k}\frac{1}{1-e^{-\beta E_\vec{k}}}
\end{equation}
We now compute $W^{(\beta)}(x-x')$:
\begin{align}
	W^{(\beta)}(x-x') &= \int \measure{p} \int\measure{q} \\
	\times &\Tr \bigg[ \rho_\beta \bigg( a_p a_q e^{-i(px + qx')}  + a_p^\dagger a_q^\dagger e^{i(px + qx')} 
    + a_p^\dagger a_q e^{-i(-px + q x')}+ a_p a_q^\dagger e^{i(-px +
qx')}\bigg)\bigg) \nonumber
\end{align}
We need
\begin{align}
		\Tr \bigg( \rho_\beta a_p^\dagger a_q \bigg)
           &= 
           \delta_{\vec{p},\vec{q}}
           \frac{
           \sum_{n_\vec{p} = 0}^\infty
		e^{-\beta E_\vec{p} n_\vec{p}} n_\vec{p}}{\sum_{n_{\vec{p}} = 0}^\infty
		e^{-\beta E_\vec{p} n_\vec{p}}
           } \,  
		=  \delta_{\vec{p},\vec{q}} \, n_\beta(E_\vec{p}) \\
		&\rightarrow (2\pi)^3 \delta^{(3)}(\vec{p} - \vec{q}) n_\beta(E_\vec{p}) \nonumber
\end{align}
where we used $\bra{\{ n_k\}} a^\dagger_p a_q \ket{ \{ n_k\}} =
\delta_{\vec{p},\vec{q}} \, n_\vec{p}$ 
and defined
the Bose-Einstein distribution factor
\begin{equation}
	n_\beta(E):= \frac{1}{e^{\beta E}-1}
  \label{eqn:nbeta}
\end{equation}
 and finally took the continuum limit in the last line. The thermal expectation
 value of $a a^\dagger$ follows from the canonical commutation relation and
 the terms with $aa, a^\dagger a^\dagger$ are zero, and thus we get
\begin{subequations}
	\begin{align}
			\Tr \bigg( \rho_\beta a_p a_q \bigg) &= 0 \\
			\Tr \bigg( \rho_\beta a_p^\dagger a_q^\dagger \bigg) &= 0 \\
			\Tr \bigg( \rho_\beta a_p^\dagger a_q \bigg) &= (2\pi)^3
			\delta^{(3)}(\vec{p} - \vec{q}) n_\beta(E_\vec{p}) \\
			\Tr \bigg( \rho_\beta a_p a_q^\dagger \bigg) &= (2\pi)^3
			\delta^{(3)}(\vec{p} - \vec{q}) \big( 1 + n_\beta(E_\vec{p}) \big)
	\end{align}
\end{subequations}
Using these, we get
\begin{subequations}
	\begin{align}
			W^{(\beta)}(x-x') &= \int \frac{\rd^3
			\vec{p}}{(2\pi)^3}\frac{1}{2E_p}\bigg[\bigg( 1 +
		n_\beta(E_\vec{p})\bigg)e^{-ip(x-x')} + n_\beta(E_\vec{p})e^{ip(x-x')} \bigg] \\
			&= \int \frac{\rd^4 p}{(2\pi)^3} \delta_+(p^2) \bigg[\bigg( 1 +
			n_\beta(-p\cdot t)\bigg)e^{-ip(x-x')} + n_\beta(-p\cdot t)e^{ip(x-x')} \bigg] \\
			&=  \int \frac{\rd^4 p}{(2\pi)^3}\bigg[\delta_+(p^2)\bigg( 1 +
			n_\beta(-p\cdot t)\bigg) + \delta_-(p^2)n_\beta(p\cdot t)\bigg] e^{-ip(x-x')}  
	\end{align}
\end{subequations}
where we changed the integration variable $p \rightarrow -p$ in the second term
of the last line. We can also show that it satisfies the KMS condition
\begin{equation}
    W^{(\beta)}(x-x') = W^{(\beta)}(x'- x + i\beta t)    
\end{equation}
which equivalently in momentum space takes the form $\tilde{W}(-p) = e^{\beta
p\cdot t} \tilde{W}(p)$. Setting $x' \rightarrow 0,\, x\rightarrow -x$, it
simplifies to $W^{(\beta)}(-x) = W^{(\beta)}(x + i\beta t)$. Writing the
left-hand side
\begin{equation}
    W^{(\beta)}(-x) = \int \frac{\rd^4 p}{(2\pi)^3} \delta_+(p^2) \bigg[\bigg( 1
			+ n_\beta(-p\cdot t)\bigg)e^{ipx} 
    + n_\beta(-p\cdot t)e^{- i p x} \bigg] 
\end{equation}
and then the right-hand side
\begin{equation}
    W^{(\beta)}(x + i\beta t) = 
    \int \frac{\rd^4 p}{(2\pi)^3} \delta_+(p^2) \bigg[n_\beta(-p\cdot t) e^{-\beta p\cdot t} e^{ipx} 
    + \bigg( 1 + n_\beta(-p\cdot t)\bigg) e^{\beta p\cdot t} e^{-ipx}  \bigg] 
\end{equation}
Equating them gives
\begin{equation}
    n_\beta(-p\cdot t) e^{-\beta p\cdot t} = 1 + n_\beta(-p\cdot t)
\end{equation}
which is satisfied given the definition of $n_\beta$ in (\ref{eqn:nbeta}).

The generalization to the gravitational field is trivially
\begin{align}
	W^{(\beta)}_{abcd}(x-x') &:= \Tr \bigg( \rho_\beta \,
	h^H_{ab}(x)h^H_{cd}\bigg) \\
													 &= \int \frac{\rd^4 p}{(2\pi)^3} \Pi_{abcd}(p) \bigg[\delta_+(p^2)\bigg( 1 +
			n_\beta(-p\cdot t)\bigg) + \delta_-(p^2)n_\beta(p\cdot t)\bigg]
			e^{-ip(x-x')}  \nonumber
\end{align}
since $\Pi_{abcd}(p)$ as defined in (\ref{eqn:polsum}) is real and even in $p$
and thus also satisfies the KMS condition.

\subsection{Squeezed state Wightman function}
\label{app:D:squeezed}
Here, we compute the Wightman function for the squeezed state
(\ref{eqn:squeezedWightman}) defined as
\begin{equation}
	W^{(\zeta)}_{abcd}(x,x') := \Tr \bigg( \rho_\zeta \, h_{ab}^H(x)h_{cd}^H(x') \bigg) 
\end{equation}
with $\rho_\zeta = \ket{\zeta}\bra{\zeta} = S(\zeta)
\ket{0}\bra{0}S^\dagger(\zeta)$ and the squeezed operator $S(\zeta)$ was defined
in (\ref{eqn:squeezeop}). The calculation amounts to evaluating the conjugation
of the fields with the squeeze operator. We first compute the conjugation of the
creation and annihilation operators. Let us define
\begin{equation}
    F(\zeta) := \frac{1}{2} \sum_s \int \frac{\rd^3 \vec{p}}{(2\pi)^3}
    \big[ 
        \zeta^{*(s)}(\vec{p}) \aa{p}{s} \aa{-p}{s} - 
        \zeta^{(s)}(\vec{p})  \ad{p}{s} \ad{-p}{s}
    \big]
\end{equation}
so that $S(\zeta) = e^{F(\zeta)}$. For the conjugation, we use the BCH formula
\begin{equation}
    e^{F(\zeta)} \aa{p}{s} e^{-F(\zeta)} = \aa{p}{s} 
    + [F(\zeta), \aa{p}{s}] 
    + \frac{1}{2!} [F(\zeta), [F(\zeta), \aa{p}{s}]] + ...
\end{equation}
which requires
\begin{subequations}
    \begin{align}
        [F(\zeta), \aa{p}{s}] &= \zeta^{(s)}(p) \ad{-p}{s} \\
        [F(\zeta), \ad{p}{s}] &= \zeta^{*(s)}(p) \aa{-p}{s} \\
        [F(\zeta), [F(\zeta), \aa{p}{s}]] 
        &= [F(\zeta), \zeta\, \ad{-p}{s}] 
        = |\zeta^{(s)}(p)|^2 \aa{p}{s}
    \end{align}
\end{subequations}
Here we used that the squeeze parameter is an even function $\zeta(p) =
\zeta(-p)$. Using these, we can sum up the series
\begin{align}
    e^{F(\zeta)} \aa{p}{s} e^{-F(\zeta)} &= 
    \aa{p}{s} 
    \bigg( 
        1 + \frac{1}{2!} |\zeta|^2 + \frac{1}{4!} |\zeta|^4 ...
    \bigg) -
    \ad{-p}{s} 
    \frac{\zeta}{|\zeta|}
    \bigg( 
        |\zeta| + \frac{1}{3!} |\zeta|^3 + \frac{1}{5!} |\zeta|^5 + ...
    \bigg) \nonumber \\
    &= \aa{p}{s}\, \cosh r^{(s)}(p) + \ad{-p}{s} e^{i\theta^{(s)}(p)} \sinh r^{(s)}(p)
\end{align}
To get $S^\dagger(\zeta)\aa{p}{s}S(\zeta) = e^{-F(\zeta)} \aa{p}{s} e^{F(\zeta)}$,
we just set $r\rightarrow -r$ so the second term flips sign
\begin{equation}
    S^\dagger(\zeta)\aa{p}{s}S(\zeta) 
    = e^{-F(\zeta)} \aa{p}{s} e^{F(\zeta)} 
    = \aa{p}{s}\, \cosh r^{(s)}(p) - \ad{-p}{s} e^{i\theta^{(s)}(p)} \sinh r^{(s)}(p)
\end{equation}
For the conjugation of $\ad{p}{s}$, we take the Hermitean conjugate of this
formula to get
\begin{equation}
    S^\dagger(\zeta)\ad{p}{s}S(\zeta) 
    = e^{-F(\zeta)} \ad{p}{s} e^{F(\zeta)} 
    = \ad{p}{s}\, \cosh r^{(s)}(p) - \aa{-p}{s} e^{-i\theta^{(s)}(p)} \sinh r^{(s)}(p)
\end{equation}
To compute the Wightman function, we need the conjugation of the product of two
mode operators such as
\begin{align}
     S^\dagger(\zeta) \aa{p}{s} \aa{q}{s'}(\zeta)  &= 
     S^\dagger(\zeta) \aa{p}{s}  S(\zeta) S^\dagger(\zeta) \aa{q}{s'} S(\zeta) \\
     &= \bigg( 
            \aa{p}{s}   \cosh r(p) - 
            \aa{-p}{s}   e^{i\theta(p)} \sinh r(p) 
        \bigg) \nonumber \\
        &\times 
        \bigg( 
            \aa{q}{s'}  \cosh r(q) - 
            \aa{-q}{s'} e^{i\theta(q)} \sinh r(q) 
        \bigg) \nonumber
\end{align}
Upon taking the vacuum expectation value, only the $aa^\dagger$ term survives
after expanding the paranthesis. The same logic applies to the rest of the
conjugations. It is convenient to introduce the quantities
\begin{equation}
    N(p) := \sinh^2 r(p), 
    \qquad
    M(p) := -\frac{e^{i\theta(p)}}{2}\sinh 2r(p),
\end{equation}
and so $\cosh^2 r(p) = 1 + N(p)$. With these, we can write the following results
\begin{subequations}
    \begin{align}
        \Tr \bigg( \rho_\zeta \aa{p}{a} \aa{q}{s'} \bigg) &= 
            (2\pi)^3 \delta_{s,s'} \delta^{(3)}(\vec{p} + \vec{q}) M(p) \\
        \Tr \bigg( \rho_\zeta \aa{p}{s} \ad{q}{s'} \bigg) &= 
            (2\pi)^3 \delta_{s,s'} \delta^{(3)}(\vec{p} - \vec{q}) \big(1+N(p)\big) \\
        \Tr \bigg( \rho_\zeta \ad{p}{s} \aa{q}{s'} \bigg) &= 
            (2\pi)^3 \delta_{s,s'} \delta^{(3)}(\vec{p} - \vec{q}) N(p) \\
        \Tr \bigg( \rho_\zeta \ad{p}{s}\ad{q}{s'} \bigg) &= 
            (2\pi)^3 \delta_{s,s'} \delta^{(3)}(\vec{p} + \vec{q}) M^*(p)
    \end{align}
\end{subequations}
Using these, the Wightman function can be written as
\begin{align}
    W^{(\zeta)}_{abcd}(x,x') &= \sum_s \int \frac{\rd^3 \vec{p}}{(2\pi)^32 E_p} 
    \bigg( \\
    & \epsilon_{ab}^{(s)}(\vec{p}) \epsilon_{cd}^{(s)}(-\vec{p}) M(p)\, e^{i E_\vec{p}(t+t')} e^{-i\vec{p}(\vec{x}-\vec{x}')} \nonumber \\
     + & \epsilon_{ab}^{(s)}(\vec{p}) \epsilon_{cd}^{*(s)}(\vec{p}) \big(1+N(p)\big)\, e^{-ip(x-x')} \nonumber \\
     + &\epsilon_{ab}^{*(s)}(\vec{p}) \epsilon_{cd}^{*(s)}(-\vec{p}) M^*(p)\, e^{-iE_\vec{p}(t+t')}e^{i\vec{p}(\vec{x}-\vec{x}')} \nonumber \\
     + &\epsilon_{ab}^{*(s)}(\vec{p}) \epsilon_{cd}^{(s)}(\vec{p}) N(p)\, e^{ip(x-x')}  \nonumber
    \bigg)
\end{align}
Recall from the beginning of section \ref{sec:3} that given our choice of
helicity basis adapted to the interferometer plane, the polarization sums in the
formula above are all equal as established in (\ref{eqn:polconj}). This allows
us to write the final covariant form
\begin{align}
    W^{(\zeta)}_{abcd}(x,x') 
			= \int \frac{\rd^4 p}{(2\pi)^3} \delta_+(p^2) \Pi_{abcd}(p)
			&\bigg[M(p) e^{ip^0(t+t')}e^{-i \vec{p}(\vec{x} - \vec{x}')} + (1+N(p))e^{-ip(x-x')} \\
            &+ M^*(p)e^{-ip^0(t+t')}e^{i \vec{p}(\vec{x} - \vec{x}')} + N(p) e^{ip(x-x')} 
			\bigg] \nonumber
\end{align}
which also correctly reduces to the vacuum in the limit $\lim_{r \rightarrow 0}
W^{(\zeta)}_{abcd} = W^{(0)}_{abcd}$.